\documentclass{article}

\usepackage{jheppub}

\usepackage{bm}
\usepackage{amsmath}
\usepackage{mathtools}
\usepackage{bbold}
\usepackage{float}
\usepackage[english]{babel}
\usepackage{caption}
\usepackage{subcaption}

\newcommand{\mm}{{\bar{m}_1}}

\def\q{\bar{q}}
\def\b{\beta^{'}}
\newcommand{\SOMMA}[2]{\displaystyle\sum\limits_{#1}^{#2}}

\newcommand{\bsigma}{{\boldsymbol \sigma}}

\newcommand{\si}{\sigma_i}

\newcommand{\sech}{\  \textnormal{sech}}

\title{About the AT line in Replica Symmetry Breaking assumptions for spin glasses}
\author[a,b]{L. Albanese} 

\affiliation[a] {Dipartimento di Matematica e Fisica ``Ennio De Giorgi'', Universit\`a del Salento, Lecce, Italy.}
\affiliation[b] {Istituto Nazionale di Fisica Nucleare, Sezione di Lecce, Italy.}

\abstract{ Replica Symmetry Breaking is a fascinating phenomenon of spin glasses model which could have consequences also in other field of studies. Although there are several studies regarding the stability between the Replica Symmetric and first step of Replica Symmetry Breaking approximations, we have very few results for the following steps (apart from that one by Gardner for P-spin glasses in 1985 and Chen in 2017 and 2021). This is link to the fact that the classic method, based from the work by De Almeida and Thoules (from which the critical stability line takes its name), is cumbersome to generalise for the next assumptions. In this paper we devise a new straightforward method inspired to the work by Toninelli in 2002 to recover the critical line in order to inspect the stability first between the second and the first steps of Replica Symmetry Breaking and then, we generalise to $K$th step, with $K$ finite.}

\begin{document}
\maketitle

\noindent
\section{Introduction}

The theory of multiple equilibria, made famous by Giorgio Parisi and his collaborators, was born in the 1970s \cite{MPV} and is still the subject of much research in the last decades in spin-glasses \cite{guerra_broken, gradenigo2020solving, GuysAlone}. The phenomenon of \textit{Replica Symmetry Breaking} (RSB), in particular, is very fascinating not only in this field, but also in very different subjects, including Artificial Intelligence (AI) \cite{Gavin-PRE2022, AABO-JPA2020, Crisanti-RSB} and biology \cite{imparato2006evaluation, peliti1994collective}. 

In a nutshell, RSB is the breaking of permutational invariance of rows and columns of order parameter matrix \cite{MPV}. Considering the associated probability distribution and regarding the finite number of steps, in the thermodynamic limit it is a sum of Dirac's deltas concentrated w.r.t. some variables between $0$ and $1$ whose summation is equal to one. Moreover, the \textit{Parisi Ansatz} allows us to show an ultrametric structure in the overlaps between the valleys \cite{ghirlanda1998general, panchenko2013parisi, talagrand2010construction}.

Already in 1978, De Almeida and Thouless were able to find a way to identify the critical line for inferring when the \textit{Replica Symmetric} (RS) approximation was stable with respect to the \textit{first step of Replica Symmetry Breaking} (1RSB) \cite{de1978stability}. This means that the probability distribution displays two peaks concentrated with respect to a variable, denoted as $\theta \in [0,1]$ in this work. De Almeida and Thouless's method consists, in a nutshell, of taking the order parameter replica matrix, adding a continuous perturbation to it and, from the stability of this perturbed matrix, we derive the critical line to be used -- called the\textit{ de Almeida-Thouless} (AT) line. Thanks to their work, several techniques are tested finding results on different models \cite{bardina2004, chen2021almeida, guerra2006replica}.\\
However, this method is very difficult to generalise to different steps  of RSB and operationally cumbersome. Moreover, it supposes that the transition is continuous, which is not true in each model as proved by Gardner for P-spin glasses \cite{gardner1985spin}. 

Toninelli in \cite{toninelli2002almeida} shows a new method to recover the AT line in SK model. It is based on the expansion of the 1RSB expression of the quenched statistical pressure w.r.t. the RS one considering that, for $\theta \in \{0,1\}$ the former returns to the latter. This method needs the knowledge of the aforementioned expression, which is not required in AT technique, however it turns out to be straightforward and rigorous and we do not suppose in advance the continuity of the transition between RS and 1RSB peaks. 
Next, this result was used by Chen in \cite{chen2021almeida, chen2017variational} and the method has been widen in \cite{albanese2023almeida} for associative neural network, in particular for Hopfield and Dense Associative Memory (DAM) models, where it has been confirmed the expression of the AT line for the Hopfield model found by Coolen in \cite{coolen2001statistical} and has been found the critical line for DAM, which was a novelty. \\
Purpose of this work is to generalize this method for the next steps of RSB for spin glasses model. In particular, we will analyze the Sherrington-Kirkpatrick (SK) model, which is the archetypal model of mean field spin glass \cite{sherrington1975solvable}, and the generalization to multi-node interactions of it, the P-spin glass model. We will see that it is possible to find a analytical critical line between two subsequent finite steps of RSB for both models. 

\par\medskip
The paper is structured as follows. \\
After the introduction of SK and P-spin glass models in Sec. \ref{sec:generalities}, we will recover in Sec. \ref{sec:appr2rsb} the expressions of the generalization of the critical lines between the second step of RSB ($2$RSB) and $1$RSB approximations for them respectevely in Subsecs. \ref{sec:skt2} and \ref{sec:P-spint2}. \\
Then, in Sec. \ref{sec:krsb}, we will generalize the procedure to $K$RSB with $K$ finite, finding also in this case the corresponding critical line both for SK and P-spin glass models.\\
Some conclusions and outlooks close the paper. Moreover, an Appendix regarding the evaluation of the correction terms in the expansion useful to recover the critical lines is reported.

\section{Generalities on Sherrington-Kirkpatrick and P-spin models}
\label{sec:generalities}
The Sherrington-Kirkpatrick (SK) model \cite{sherrington1975solvable} is a system of $N$ Ising spins $\sigma_i \in \{ \pm 1 \}$ interacting via symmetric pairwise interactions 
$J_{ij}$ which are i.i.d. Gaussian variables with zero average and variance $J^2/N$. The Hamiltonian of the model is 
\begin{align}
    H_N(\bm \sigma \vert J) := - \dfrac{1}{2} \sum_{i,j}^{N,N} J_{ij} \si \sigma_j
\end{align}
and the order parameter is the two-replica overlap {$q$} 
\begin{eqnarray}
\label{eq:overlap}
q(\bsigma^{(1)},\bsigma^{(2)})\!\!:=\dfrac{1}{N}\SOMMA{i=1}{N}\sigma_i^{(1)}\sigma_i^{(2)},
\end{eqnarray}
which quantifies the correlations between two 
configurations $\bsigma^{(1)},\bsigma^{(2)}$ of the system, with the same realization of the patterns (i.e. quenched disorder).

Then, we denote the associated Boltzmann factor, at inverse temperature $\beta=1/T$, as
\begin{equation}
{B}_N(\bm{\sigma}|\bm{J})=\frac{e^{-\beta H_N(\bm \sigma \vert \bm J )}}{Z_N(\bm J)},\quad\quad\quad Z_N(\bm J)=\sum_{\bsigma}e^{-\beta H_N(\bm \sigma \vert \bm J )},
\end{equation}
where $Z_N(\bm J)$ is the corresponding partition function. 

Rather than considering pair-wise interactions, we can move to systems with $P>2$ even multi-node interactions, known as P-spin. In this case, the Hamiltonian of the model is stated as 
\begin{align}
    H_N(\bm \sigma\vert J) &= -\dfrac{1}{N^{P-1}P!} \sum_{i_1, \hdots , i_P=1, \hdots , 1}^{N, \hdots , N} J_{i_1, \hdots, i_P} \sigma_{i_1} \hdots \sigma_{i_P}
    \label{eq:HamiltonianP-spin}
\end{align}
where $J_{i_1, \hdots, i_P}$ are Gaussian i.i.d. variables, $J_{i_1, \hdots, i_P} \sim \mathcal{N}(0,J^2)$.
The order parameter of the network is still the overlap between two replicas defined in \eqref{eq:overlap}.

Last but not least, we introduce the quenched statistical pressure as 
\begin{align}
    \mathcal{A}(\beta)= \lim_{N \to +\infty} \mathcal{A}_N(\beta) = \lim_{N \to +\infty} \dfrac{1}{N} \mathbb{E} \log Z_N(\bm J)=-\lim_{N \to +\infty}\beta f_N(\beta)=-\beta f(\beta),
\end{align}
where $\mathbb{E}$ is the expectation w.r.t. all the i.i.d. $J_{ij}$ gaussian standard variables, whereas $f_N$ is known as the quenched free energy. Generally, one can use either quenched statistical pressure or free energy to do the computations. The existence and uniqueness of the quenched free energy is proven for both SK and Pspin model in \cite{guerra2002thermodynamic, talagrand2006parisi}.

\section{Critical line between 2RSB and 1RSB approximations}
\label{sec:appr2rsb}
Purpose of this Section is to find the expression of the critical line which generalizes the AT line in RSB steps for the two models introduced in Sec. \ref{sec:generalities}. We will use an approach inspired to that one by Toninelli \cite{toninelli2002almeida}. In particular, we are now interested in the stability of 2RSB assumption with respect to the 1RSB one.

\subsection{SK model}
\label{sec:skt2}
1RSB scheme 
\cite{Crisanti-RSB,Steffan-RSB,AABO-JPA2020} assumes that
  the distribution of the two-replica overlap $q$, in the thermodynamic limit, displays two delta-peaks at the equilibrium values $\tilde q_0$ and $\tilde q_1>\tilde q_0$ and the concentration on these two values is ruled by the parameter $\theta \in [0, 1]$: 
\begin{align}
    \lim_{N \rightarrow + \infty} P'_N(q) &= \theta \delta (q - \tilde{q}_0) + (1-\theta) \delta (q - \tilde{q}_1).    \label{eq:1RSBAss} 
\end{align}  
We stress that either for $\theta=0$ or for $\theta=1$ the probability distribution reduces to a delta concentrated on a single peak (which is known as RS assumption). 

The 1RSB approximation of the quenched free-energy is (see e.g. \cite{nishimori2001statistical})
\begin{align}
    -\beta f_{1RSB}(\q_1, \q_0 \vert \beta, J, \theta) =& \ln 2 + \dfrac{\beta^2 J^2}{2} (1-\tilde q_1)+\dfrac{1}{\theta} \mathbb{E}_1 \left[ \ln \mathbb{E}_2 \cosh^\theta \tilde g(\tilde q_1, \tilde q_0 \vert \beta, J )\right]\notag \\
        &-\dfrac{\beta^2 J^2}{4} \left[1-\tilde q_1^2 +\theta (\tilde q_1^2-\tilde q_0^2) \right],
    \label{eq:A_1RSB}
\end{align}
where 
\begin{align}
\tilde g(\tilde q_1, \tilde q_0 \vert \beta, J)=\beta J \sqrt{\tilde q_1-\tilde q_0} z^{(2)} + \beta J \sqrt{\tilde q_0} z^{(1)}
\label{eq:gSK}
\end{align}
and 
$\mathbb{E}_1$, $\mathbb{E}_2$ are the average w.r.t. the standard Gaussian variables $z^{(1)}$ and $z^{(2)}$, respectively. The order parameters $\tilde{q}_0$ and  $\tilde{q}_1$ fulfill the following self-consistency equations
\begin{align}
    \tilde q_1 &= \mathbb{E}_1 \left[ \dfrac{\mathbb{E}_2 \cosh^\theta \tilde g(\tilde q_1, \tilde q_0 \vert \beta, J) \tanh^2 \tilde g(\tilde q_1, \tilde q_0\vert \beta, J)}{\mathbb{E}_2 \cosh^\theta \tilde g(\tilde q_1, \tilde q_0\vert \beta, J)}\right], \notag \\
    \tilde q_0 &= \mathbb{E}_1 \left[ \dfrac{\mathbb{E}_2 \cosh^\theta \tilde g(\tilde q_1, \tilde q_0\vert \beta, J) \tanh \tilde g(\tilde q_1, \tilde q_0\vert \beta, J)}{\mathbb{E}_2 \cosh^\theta \tilde g(\tilde q_1, \tilde q_0\vert \beta, J)}\right]^2.
    \label{eq:self1RSB_SK}
\end{align}

In 2RSB approximation the distribution of the overlap $q$ displays three delta-peaks at the equilibrium values $\q_0$, $\q_1 > \q_0$ and $\q_2>\q_1$ concentrated w.r.t. two parameters $\theta_1 \in [0,1]$ and $\theta_2 \in [0,1]$:
\begin{align}
    \lim_{N \to + \infty} P''_N(q)=\theta_1 \delta(q-\q_0)+(\theta_2-\theta_1) \delta(q-\q_1) + (1-\theta_2) \delta(q-\q_2).
    \label{2RSBass}
\end{align}

We note that for $\theta_1 =0$, $\theta_2=\theta_1$ and $\theta_2=1$ we come back to the 1RSB assumption. 

The 2RSB approximation of the quenched statistical pressure, instead, is 
\begin{align}
     -\beta f_{2RSB}(\q_2, \q_1, \q_0 \vert \beta, J, \theta_2, \theta_1) =& \ln 2 + \dfrac{\beta^2 J^2}{4} \left[ 1-(\theta_2-1) \q_2^2 + (\theta_2-\theta_1) \q_1^2 + \theta_1 \q_0^2 \right] - \dfrac{\beta^2 J^2}{2} \q_2 \notag \\
     &+ \dfrac{1}{\theta_1} \mathbb{E}_1 \log \mathbb{E}_2 \left[ \mathbb{E}_3 \cosh^{\theta_2} g(\q_2, \q_1, \q_0\vert \beta, J) \right]^{\theta_1/\theta_2}
\end{align}
where 
\begin{align}
    g(\q_2, \q_1, \q_0\vert \beta, J) = \beta J\sqrt{\q_0} z^{(1)} + \beta J\sqrt{\q_1-\q_0} z^{(2)} + \beta J\sqrt{\q_2-\q_1} z^{(3)}
\end{align}
and $\mathbb{E}_a$ is the average w.r.t. $z^{(a)}$, for $a=1,2,3$.

From now on, we imply the dependence of $g$, $f_{1RSB}$ and $f_{2RSB}$ on $\beta$ and $J$. 
The order parameters $\q_0, \ \q_1, \ \q_2$ fulfill the following self-consistency equations 
\begin{align}
    \q_0=& \mathbb{E}_1 \left\{ \dfrac{\mathbb{E}_2 \left[\left( \mathbb{E}_3 \cosh^{\theta_2} g(\q_2,\q_1, \q_0)\right)^{\theta_1/\theta_2} \dfrac{\mathbb{E}_3 \cosh^{\theta_2}g(\q_2, \q_1, \q_0) \tanh g(\q_2, \q_1, \q_0) }{\mathbb{E}_3 \cosh^{\theta_2}g(\q_2, \q_1, \q_0)}\right]}{\mathbb{E}_2 \left[ \mathbb{E}_3 \cosh^{\theta_2} g(\q_2,\q_1, \q_0)\right]^{\theta_1/\theta_2}}\right\}^2, \label{eq:q0_2RSB}\\
    \q_1=& \mathbb{E}_1 \left\{ \dfrac{\mathbb{E}_2 \left[\left( \mathbb{E}_3 \cosh^{\theta_2} g(\q_2,\q_1, \q_0)\right)^{\theta_1/\theta_2} \left(\dfrac{\mathbb{E}_3 \cosh^{\theta_2}g(\q_2, \q_1, \q_0) \tanh g(\q_2, \q_1, \q_0) }{\mathbb{E}_3 \cosh^{\theta_2}g(\q_2, \q_1, \q_0)}\right)^2\right]}{\mathbb{E}_2 \left[ \mathbb{E}_3 \cosh^{\theta_2} g(\q_2,\q_1, \q_0)\right]^{\theta_1/\theta_2}}\right\}, \\
    \q_2=& \mathbb{E}_1 \left\{ \dfrac{\mathbb{E}_2 \left[\left( \mathbb{E}_3 \cosh^{\theta_2} g(\q_2,\q_1, \q_0)\right)^{\theta_1/\theta_2} \dfrac{\mathbb{E}_3 \cosh^{\theta_2}g(\q_2, \q_1, \q_0) \tanh^2 g(\q_2, \q_1, \q_0) }{\mathbb{E}_3 \cosh^{\theta_2}g(\q_2, \q_1, \q_0)}\right]}{\mathbb{E}_2 \left[ \mathbb{E}_3 \cosh^{\theta_2} g(\q_2,\q_1, \q_0)\right]^{\theta_1/\theta_2}}\right\}. \label{eq:q2_2RSB}
\end{align}

In order to apply the method explained in \cite{albanese2023almeida}, we need to decide 
which limit of the probability distribution to consider to go from the 2RSB probability distribution to the 1RSB one. 

We will see in details now the $\theta_2\ \to 1$ case and we will report only the results for the $\theta_1 \to 0$ and $\theta_2 \to \theta_1$ limits.

\par\medskip
Our purpose is then to prove that for values of $\theta_2$ close but away from one, the 2RSB expression of the quenched 
free-energy is 
smaller
than the 1RSB expression, i.e. 
$f_{2RSB}(\bar{q}_2, \bar{q}_{1},\bar{q}_0|\theta)<f_{1RSB}(\tilde{q}_1, \tilde q_0)$
below a critical line in the parameters space $(J,\beta)$.  

To this purpose, we expand the 2RSB quenched 
free-energy 
around $\theta_2=1$ (i.e. around the 1RSB expression) to the first order, writing
\begin{align} 
   f_{2RSB} (\bar{q}_2, \bar{q}_1,\bar{q}_0 \vert \theta)=& f_{2RSB}( \bar{q}_2, \bar{q}_1,\bar{q}_0 \vert \theta_2)|_{\theta_2=1}
   + (\theta_2-1) \partial_{\theta_2} f_{1RSB}(\bar{q}_2, \bar{q}_1,\bar{q}_0 \vert \theta_2)\vert_{\theta_2=1},
    \label{eq:expansionSK_t1}
\end{align}
where $f_{2RSB}(\bar{q}_2, \bar{q}_1,\bar{q}_0\vert \theta_2)|_{\theta_2=1}= f_{1RSB}(\bar{q}_1, \bar{q}_0)$. To determine when the RS solution becomes unstable, i.e.  $f_{2RSB}(\bar{q}_2, \bar{q}_1,\bar{q}_0\vert \theta_2)<f_{1RSB}(\bar{q}_1, \bar{q}_0)$, we inspect the sign of $\partial_{\theta_2} f_{2RSB}(\bar{q}_2, \bar{q}_1,\bar{q}_0\vert \theta)\vert_{\theta_2=1}$, keeping in mind that $\theta_2-1 <0$.

To evaluate the latter, we need to expand the self-consistency equations for $\q_0$, $\q_1$ and $\q_2$ around $\theta_2=1$, to linear orders in $\theta_2-1$. 
We obtain

\begin{align}
    \q_0 &= \mathbb{E}_1 \left\{ \dfrac{\mathbb{E}_2 \cosh^{\theta_1} \tilde g( \q_1, \q_0)\tanh \tilde g(\q_1, \q_0) } {\mathbb{E}_2 \cosh^{\theta_1} \tilde g( \q_1, \q_0)}\right\}^2 + (\theta_2-1) C(\q_0, \q_1, \q_2),
    \label{eq:q0exp_SK}
\end{align}
where $C(\q_0, \q_1, \q_2)$ is reported in \eqref{eq:C}. Specifically, we notice that we can rewrite \eqref{eq:q0exp_SK} as 
\begin{align}
    \q_0=\tilde q_0 + (\theta_2-1) C(\q_0, \q_1, \q_2).
    \label{q0_expSK}
\end{align}

Moving on to $\q_1$, we have 
\begin{align}
    \q_1 = \mathbb{E}_1 \left\{ \dfrac{\mathbb{E}_2 \left[\cosh^{\theta_1} \tilde g(\q_1, \q_0)\tanh \tilde g( \q_1, \q_0)\right]^2 } {\mathbb{E}_2 \cosh^{\theta_1} \tilde g(\q_1, \q_0)}\right\} + (\theta_2-1) B(\q_0, \q_1, \q_2), 
\end{align}
with $B(\q_0, \q_1, \q_2)$ in \eqref{eq:B} and we have also in this case
\begin{align}
    \q_1 = \tilde q_1 + (\theta_2-1) B(\q_0, \q_1, \q_2).
    \label{q1_expSK}
\end{align}
Then, 
\begin{align}
    \q_2 =& \mathbb{E}_1 \left\{ \dfrac{\mathbb{E}_2 \cosh^{\theta_1} \tilde g( \q_1, \q_0)\dfrac{\mathbb{E}_3 \cosh g (\q_0, \q_1, \q_2) \tanh^2 g (\q_0, \q_1, \q_2, \mm)}{\mathbb{E}_3 \cosh^{\theta_1} g (\q_0, \q_1, \q_2)}} {\mathbb{E}_2 \cosh^{\theta_1} \tilde g(\mm, \q_1, \q_0)}\right\} \notag \\
    &+ (\theta_2-1) A(\q_0, \q_1, \q_2),
\end{align}
where $A(\q_0, \q_1, \q_2)$ is reported in \eqref{eq:A}. We denote 
\begin{align}
    \tilde q_2(\tilde q_1, \tilde q_0) = \mathbb{E}_1 \left\{ \dfrac{\mathbb{E}_2 \cosh^{\theta_1} \tilde g(\q_1, \q_0)\dfrac{\mathbb{E}_3 \cosh g (\q_0, \q_1, \q_2 ) \tanh^2 g (\q_0, \q_1, \q_2)}{\mathbb{E}_3 \cosh^{\theta_1} g (\q_0, \q_1, \q_2)}} {\mathbb{E}_2 \cosh^{\theta_1} \tilde g(\q_1, \q_0)}\right\}
    \label{eq:q2t1_SK}
\end{align}
which does not have correspondence on the 1RSB peaks. Therefore we can rewrite the expression for $\q_2$ as 
\begin{align}
    \q_2 = \tilde q_2(\tilde q_1, \tilde q_0) + (\theta_2-1) A(\q_0, \q_1, \q_2).
    \label{q2_expSK}
\end{align}

Using  \eqref{q0_expSK},\eqref{q1_expSK}, \eqref{q2_expSK} we can rewrite the expression of the quenched free energy as 
\begin{align}
     -\beta f_{2RSB}&(\tilde q_2, \tilde q_1, \tilde q_0 \vert \theta_2, \theta_1)=\ln 2 + \dfrac{1}{\theta_1} \mathbb{E}_1 \ln \mathbb{E}_2 \left[ \mathbb{E}_3 \cosh^{\theta_2} g (\tilde q_0, \tilde q_1, \tilde q_2)\right]^{\theta_1/\theta_2} \notag \\
     &+ \dfrac{\beta^2 J^2}{4} \left[ 1- (\theta_2 - 1) \tilde q_2^2 + (\theta_2-1) \tilde q_1^2 + (1-\theta_1) \left( \tilde q_1^2 + 2 \tilde q_1 (\theta_2-1) B(\q_0, \q_1, \q_2)\right) \right] \notag \\
     &+ \dfrac{\beta^2 J^2}{4} \left[ \theta_1 \left( \tilde q_0^2 + 2 (\theta_2-1) \tilde q_0 C(\q_0, \q_1, \q_2)\right)\right] - \dfrac{\beta^2 J^2}{2} \left( \tilde q_2 + (\theta_2-1) A(\q_0, \q_1, \q_2)\right). 
\end{align}

Computing the derivative we need in the expansion \eqref{eq:expansionSK_t1} we get 
\begin{align}
    K(\tilde q_0, \tilde q_1, \tilde q_2(\tilde q_1, \tilde q_0)) =& \dfrac{\beta^2 J^2}{4} \left( \tilde q_1^2 - \tilde q_2^2 \right) - \mathbb{E}_1 \dfrac{\mathbb{E}_2 \left[ \left( \mathbb{E}_3 \cosh g(\tilde q_0, \tilde q_1, \tilde q_2) \right)^{\theta_1} \ln \mathbb{E}_3 \cosh g(\tilde q_0, \tilde q_1, \tilde q_2) \right]}{\mathbb{E}_2 \left[ \mathbb{E}_3 \cosh g(\tilde q_0, \tilde q_1, \tilde q_2)\right]^{\theta_1}} \notag \\
    &+ \mathbb{E}_1 \dfrac{\mathbb{E}_2 \left[ \left( \mathbb{E}_3 \cosh g(\tilde q_0, \tilde q_1, \tilde q_2) \right)^{\theta_1}\dfrac{\mathbb{E}_3 \cosh g(\tilde q_0, \tilde q_1, \tilde q_2) \ln \cosh g(\tilde q_0, \tilde q_1, \tilde q_2) }{\mathbb{E}_3 \cosh g(\tilde q_0, \tilde q_1, \tilde q_2)}\right]}{\mathbb{E}_2 \left[ \mathbb{E}_3 \cosh g(\tilde q_0, \tilde q_1, \tilde q_2)\right]^{\theta_1}}. 
    \label{K_SK}
\end{align}
Therefore, we need to study the sign of \eqref{K_SK}, where $\tilde q_0, \tilde q_1$ and $\tilde q_2(\tilde q_1, \tilde q_0)$ are the solutions of the self-consistency equations \eqref{eq:self1RSB_SK} and \eqref{eq:q2t1_SK}. To do so, we are going to study the behaviour of the function $ K(\tilde q_0, \tilde q_1, x)$ for $x \in [0, \tilde q_1]$. We notice that for $x=\tilde q_1$ we get $K(\tilde q_0, \tilde q_1, \tilde q_1)=0$. The extremum of the function, supposing it is the global one, is 
\begin{align}
    \partial_x  K(\tilde q_0, \tilde q_1, x) = - \dfrac{\beta^2 J^2}{2} \left\{ x- \mathbb{E}_1 \left[\dfrac{\mathbb{E}_2 \cosh^{\theta_1} \tilde g(\tilde q_1, \tilde q_0)\dfrac{\mathbb{E}_3 \cosh g (\q_0, \q_1, x ) \tanh^2 g (\q_0, \q_1, x)}{\mathbb{E}_3 \cosh^{\theta_1} g (\q_0, \q_1, \q_2)}} {\mathbb{E}_2 \cosh^{\theta_1} \tilde g(\q_1, \q_0)}\right]\right\}
\end{align}
which is nullified when $x$ is exactly $\tilde q_2(\tilde q_0, \tilde q_1)$: 
\begin{align}
    x= \mathbb{E}_1 \left\{ \dfrac{\mathbb{E}_2 \cosh^{\theta_1} \tilde g(\q_1, \q_0)\dfrac{\mathbb{E}_3 \cosh g (\q_0, \q_1, x ) \tanh^2 g (\q_0, \q_1, x)}{\mathbb{E}_3 \cosh^{\theta_1} g (\q_0, \q_1, x)}} {\mathbb{E}_2 \cosh^{\theta_1} \tilde g(\q_1, \q_0)}\right\} \equiv \tilde q_2(\tilde q_0, \tilde q_1).
\end{align}

Considering that $ K(\tilde q_0, \tilde q_1, x)$ vanishes for $x=\tilde q_1$ and we have supposed that the extremum $x= \tilde q_2(\tilde q_0, \tilde q_1)$ is global in the domain considered, we have that $ K(\tilde q_0, \tilde q_1, x) >0 $ if $x$ is a maximum and $ K(\tilde q_0, \tilde q_1, x)$ is $x$ is a minimum. Therefore, if 

\begin{align}
    \partial_{x^2}  K(\tilde q_0, \tilde q_1, x) \Big\vert_{x= \tilde q_2(\tilde q_0, \tilde q_1)} = - \dfrac{\beta^2 J^2}{2} \left\{ 1-\beta^2 J^2\mathbb{E}_1 \left[ \dfrac{\mathbb{E}_2 \cosh^{\theta_1}  \tilde g(\tilde q_1, \tilde q_0) \dfrac{\mathbb{E}_3 \sech^3 g (\q_0, \q_1, \tilde q_2(\tilde q_0, \tilde q_1) ) }{\mathbb{E}_3 \cosh g (\q_0, \q_1, \tilde q_2(\tilde q_0, \tilde q_1) ) }}{\mathbb{E}_2 \left( \cosh^{\theta_1}  \tilde g(\tilde q_1, \tilde q_0)\right)}\right]\right\}
\end{align}
is positive, $ K(\tilde q_0, \tilde q_1, x)$ is negative and 
\begin{align}
    f_{2RSB} (\bar{q}_2, \bar{q}_1,\bar{q}_0 \vert \theta)=& f_{2RSB}( \bar{q}_2, \bar{q}_1,\bar{q}_0 \vert \theta_2)|_{\theta_2=1}
   + (\theta_2-1) K(\tilde q_0, \tilde q_1, \tilde q_2(\tilde q_0, \tilde q_1)) < f_{1RSB}(\bar{q}_1, \bar{q_0}).
\end{align}
Hence the $1RSB$ assumption becomes unstable when 
\begin{align}
    1-\beta^2 J^2 \mathbb{E}_1 \left[ \dfrac{\mathbb{E}_2 \cosh^{\theta_1}  \tilde g(\tilde q_1, \tilde q_0) \dfrac{\mathbb{E}_3 \sech^3 g (\q_0, \q_1, \tilde q_2(\tilde q_0, \tilde q_1) ) }{\mathbb{E}_3 \cosh g (\q_0, \q_1, \tilde q_2(\tilde q_0, \tilde q_1) ) }}{\mathbb{E}_2 \left( \cosh^{\theta_1}  \tilde g(\tilde q_1, \tilde q_0)\right)}\right] <0.
    \label{eq:lineat2}
\end{align}
Considering the results from \cite{albanese2023almeida}, if we put $\tilde q_2 (\tilde q_1, \tilde q_0) \to \tilde q_1$ we recover the corresponding of the AT line between RS and 1RSB assumptions:
\begin{align}
        1-\beta^2 J^2\mathbb{E}_1 \left[ \dfrac{\mathbb{E}_2 \cosh^{\theta_1-4}  \tilde g(\tilde q_1, \tilde q_0)}{\mathbb{E}_2 \cosh^{\theta_1}  \tilde g(\tilde q_1, \tilde q_0)}\right]<0,
        \label{eq:AT2RSB_SK}
\end{align}
which corresponds to the critical line found by \cite{chen2017variational, gardner1985spin}.

We stress that, for $\theta_1\to \{ 0, 1\}$ in \eqref{eq:lineat2} and $\tilde q_2 (\tilde q_1, \tilde q_0) \to \tilde q_1$  we recover the classic AT line \cite{de1978stability}:
\begin{align}
    1-\beta^2J^2 \mathbb{E} \textrm{sech}^4\left( \beta J \sqrt{\q} z \right) <0.
\end{align}

Using the same procedure we can compute the corresponding critical line in $\theta_1 \to 0$ and $\theta_2 \to \theta_1$ limits. 

For the former, we get 
\begin{align}
    1-\beta^2 J^2 &\mathbb{E}_1 \left\{ \mathbb{E}_2 \left[ \dfrac{\mathbb{E}_3 \cosh^{\theta_2}\tilde g (\tilde q_1, \tilde q_0, \tilde q_2(\tilde q_0, \tilde q_1) )\textrm{sech}^2 g(\tilde q_1, \tilde q_0, \tilde q_2(\tilde q_0, \tilde q_1) ) }{\mathbb{E}_3 \cosh^{\theta_2} g(\tilde q_1, \tilde q_0, \tilde q_2(\tilde q_0, \tilde q_1) )}\right]\right. \notag \\
    &+ \theta_2 \left[  \mathbb{E}_2 \left( \dfrac{\mathbb{E}_3 \cosh^{\theta_2} g(\tilde q_1, \tilde q_0, \tilde q_2(\tilde q_0, \tilde q_1) ) \tanh^2 g(\tilde q_1, \tilde q_0, \tilde q_2(\tilde q_0, \tilde q_1) )}{\mathbb{E}_3 \cosh^{\theta_2} g(\tilde q_1, \tilde q_0, \tilde q_2(\tilde q_0, \tilde q_1) ) } \right) \right. \notag \\
    &\left. \left.- \mathbb{E}_2 \left( \dfrac{\mathbb{E}_3 \cosh^{\theta_2} g(\tilde q_1, \tilde q_0, \tilde q_2(\tilde q_0, \tilde q_1) ) \tanh g(\tilde q_1, \tilde q_0, \tilde q_2(\tilde q_0, \tilde q_1) )}{\mathbb{E}_3 \cosh^{\theta_2} g(\tilde q_1, \tilde q_0, \tilde q_2(\tilde q_0, \tilde q_1) ) } \right)^2 \right]\right\}^2 <0,
\end{align}
where $g(\tilde q_1, \tilde q_0, \tilde q_2(\tilde q_0, \tilde q_1) )$ is defined as 
\begin{align}
    g(\tilde q_1, \tilde q_0, \tilde q_2(\tilde q_0, \tilde q_1) ) = \beta J \sqrt{\tilde q_2(\tilde q_0, \tilde q_1) } z^{(1)} + \beta J \sqrt{\tilde q_0 - \tilde q_2 (\tilde q_0, \tilde q_1) } z^{(2)} + \beta J \sqrt{\tilde q_1 - \tilde q_0} z^{(3)}
    \end{align}
    where $\tilde q_1$ and $\tilde q_0$ are the solutions of \eqref{eq:self1RSB_SK} and $\tilde q_2$ fulfills the following 
    \begin{align}
        \tilde q_2(\tilde q_0, \tilde q_1) = \mathbb{E}_1 \left\{ {\mathbb{E}_2 \left[ \dfrac{\mathbb{E}_3 \cosh^{\theta_2} g(\tilde q_1, \tilde q_0, \tilde q_2(\tilde q_0, \tilde q_1) ) \tanh g(\tilde q_1, \tilde q_0, \tilde q_2(\tilde q_0, \tilde q_1) )}{\mathbb{E}_3 \cosh^{\theta_2} g(\tilde q_1, \tilde q_0, \tilde q_2(\tilde q_0, \tilde q_1) )}\right]}\right\}^2.
    \end{align}

If $\tilde q_2 (\tilde q_0, \tilde q_1) \to \tilde q_0$ we get an expression which is cumbersome w.r.t. \eqref{eq:AT2RSB_SK}: 
\begin{align}
    1-\beta^2 J^2 &\mathbb{E}_1 \left\{ \dfrac{\mathbb{E}_3 \cosh^{\theta_2}\tilde g (\tilde q_1, \tilde q_0 )\textrm{sech}^2 g(\tilde q_1, \tilde q_0 ) }{\mathbb{E}_3 \cosh^{\theta_2} g(\tilde q_1, \tilde q_0 )}\right. \notag \\
    &\left.+ \theta_2 \left[   \left( \dfrac{\mathbb{E}_3 \cosh^{\theta_2} g(\tilde q_1, \tilde q_0 ) \tanh^2 g(\tilde q_1, \tilde q_0 )}{\mathbb{E}_3 \cosh^{\theta_2} g(\tilde q_1, \tilde q_0 ) } \right) -  \left( \dfrac{\mathbb{E}_3 \cosh^{\theta_2} g(\tilde q_1, \tilde q_0 ) \tanh g(\tilde q_1, \tilde q_0 )}{\mathbb{E}_3 \cosh^{\theta_2} g(\tilde q_1, \tilde q_0 ) } \right)^2 \right] \right\}^2 <0.
    \label{eq:lineat1}
\end{align}

However, for $\theta_2 \to \{ 0, 1\}$ in \eqref{eq:lineat1} we get exactly the AT line expression. This assure the coherence of the computations. Both Eqs. \eqref{eq:lineat2} and \eqref{eq:lineat1} return exactly to the 1RSB expressions found in \cite{albanese2023almeida}.

In $\theta_2 \to \theta_1$ limit, instead, we have\footnote{We stress that the $\theta_1 \to \theta_2$ limit takes us to the same results.}
\begin{align}
1-\beta^2 J^2 &\mathbb{E}_1 \left\{ \dfrac{1}{\mathbb{E}_2 \mathbb{E}_3 \cosh^{\theta_1} g(\tilde q_1, \tilde q_0, \tilde q_2(\tilde q_0, \tilde q_1) )}\mathbb{E}_2 \left\{ \dfrac{1}{\mathbb{E}_3 \cosh^{\theta_1} g(\tilde q_1, \tilde q_0, \tilde q_2(\tilde q_0, \tilde q_1) )}\right.\right.\notag \\
    &\cdot \left[ \mathbb{E}_3 \cosh^{\theta_1}g(\tilde q_1, \tilde q_0, \tilde q_2(\tilde q_0, \tilde q_1) )\sech^2 g(\tilde q_1, \tilde q_0, \tilde q_2(\tilde q_0, \tilde q_1) ) \right. \notag \\
    &+ \theta_1 \left( \mathbb{E}_3 \cosh^{\theta_1} g(\tilde q_1, \tilde q_0, \tilde q_2(\tilde q_0, \tilde q_1) ) \tanh^2 g(\tilde q_1, \tilde q_0, \tilde q_2(\tilde q_0, \tilde q_1) ) \right.\notag \\
&\left.\left.\left.\left.- \dfrac{\left( \mathbb{E}_3 \cosh^{\theta_1} g(\tilde q_1, \tilde q_0, \tilde q_2(\tilde q_0, \tilde q_1) ) \tanh g(\tilde q_1, \tilde q_0, \tilde q_2(\tilde q_0, \tilde q_1) )\right)^2}{\mathbb{E}_3 \cosh^{\theta_1}g(\tilde q_1, \tilde q_0, \tilde q_2(\tilde q_0, \tilde q_1) )}\right) \right]^2\right\}\right\} <0,
\end{align}
where $g(\tilde q_1, \tilde q_0, \tilde q_2(\tilde q_0, \tilde q_1) )$ is defined as 
\begin{align}
    g(\tilde q_1, \tilde q_0, \tilde q_2(\tilde q_0, \tilde q_1) ) = \beta J \sqrt{\tilde q_0} z^{(1)} + \beta J \sqrt{\tilde q_{2}(\tilde q_0, \tilde q_1) - \tilde q_0} z^{(2)} + \beta J \sqrt{\tilde q_1 - \tilde q_2(\tilde q_0, \tilde q_1)}z^{(3)}
\end{align}
and $\tilde q_0, \ \tilde q_1$ are always the solutions of \eqref{eq:self1RSB_SK} and, instead, $\tilde q_2$ fulfills the following equation
\begin{align}
    \tilde q_2 (\tilde q_0, \tilde q_1) = \mathbb{E}_1 \left\{ \dfrac{\mathbb{E}_2 \left[ \dfrac{\left(\mathbb{E}_3 \cosh^{\theta_1} g(\tilde q_1, \tilde q_0, \tilde q_2(\tilde q_0, \tilde q_1) )\tanh g(\tilde q_1, \tilde q_0, \tilde q_2(\tilde q_0, \tilde q_1) )\right)^2}{\mathbb{E}_3 \cosh^{\theta_1} g(\tilde q_1, \tilde q_0, \tilde q_2(\tilde q_0, \tilde q_1) )}\right]}{\mathbb{E}_2\mathbb{E}_3 \cosh^{\theta_1} g(\tilde q_1, \tilde q_0, \tilde q_2(\tilde q_0, \tilde q_1) )}\right\}.
\end{align}
In this case, if $\tilde q_2 (\tilde q_0, \tilde q_1) \to \tilde q_0$ we get 
\begin{align}
    1-\beta^2 J^2 &\mathbb{E}_1 \left\{ \dfrac{1}{\mathbb{E}_2 \cosh^{\theta_1} g(\tilde q_0, \tilde q_1)} \mathbb{E}_2 \left\{ \dfrac{1}{ \cosh^{\theta_1}g(\tilde q_0, \tilde q_1)}\left[ \cosh^{\theta_1} g(\tilde q_0, \tilde q_1) \textrm{sech}^2 g(\tilde q_0, \tilde q_1) \right.\right.\right.\notag \\
    &\left. \left. \left.+ \theta_1 \left( \cosh^{\theta_1} g(\tilde q_0, \tilde q_1) \tanh^2 g(\tilde q_0, \tilde q_1) - \cosh^{\theta_1} g(\tilde q_0, \tilde q_1) \tanh^2 g(\tilde q_0, \tilde q_1)\right)\right]^2\right\}\right\} <0,
\end{align}
which becomes the AT line when $\theta_1 \to \{0,1\}. $

Thus, we highlight that the three limits create three different equations which collapse analytically only in the RS case (the already known AT line). 


\subsection{P-spin glass model}
\label{sec:P-spint2}
Now we analyze the situation of the P-spin glass model, which we recall it is a generalization of mean-field spin glass with a strength of interactions which is greater than $2$. \\
The order parameter is the same in \eqref{eq:overlap} and in 1RSB approximation it behaves in the same way as in SK model \eqref{eq:1RSBAss}. 

The 1RSB expression of the quenched free energy in the thermodynamic limit is \cite{gardner1985spin}
\begin{align}
    -\b f_{1RSB}(\tilde q_1, \tilde q_0 \vert \beta', J, \theta) =& \ln 2 + \dfrac{{\b}^2 J^2}{4} \left[ 1-P\tilde q_1^{P-1} + (P-1) \tilde q_1^P \right] \notag \\
    &+ \dfrac{1}{\theta} \mathbb{E}_1 \ln \mathbb{E}_2 \cosh^\theta \tilde g (\beta', J, \tilde q_1, \tilde q_0)  - \dfrac{{\b}^2 J^2}{4} (P-1)\theta (\tilde q_1^P-\tilde q_0^P)
    \label{eq:ATSKK}
\end{align}
where $\beta ' = \dfrac{2\beta}{P!} $, $\mathbb{E}_1$, $\mathbb{E}_2$ are the average w.r.t. the standard Gaussian variables $z^{(1)}$ and $z^{(2)}$
\begin{align}
     \tilde g(\beta', J, \tilde q_1, \tilde q_0) =\b J z^{(1)} \sqrt{\dfrac{P}{2} \tilde q_0^{P-1}} + \b J z^{(2)} \sqrt{\dfrac{P}{2} (\tilde q_1^{P-1}-\tilde q_0^{P-1})},
     \label{eq:gPSK}
\end{align}
 and $\tilde q_0, \ \tilde q_1$ fulfill the following self-consistency equations 
\begin{equation}
\begin{array}{lll}
    \tilde q_1 &= \mathbb{E}_1 \left[ \dfrac{\mathbb{E}_2 \cosh^\theta \tilde g(\beta', J, \tilde q_1, \tilde q_0) \tanh^2 \tilde g(\beta', J, \tilde q_1, \tilde q_0)}{\mathbb{E}_2 \cosh^\theta \tilde g(\beta', J, \tilde q_1, \tilde q_0)}\right], \\\\
    \tilde q_0 &= \mathbb{E}_1 \left[ \dfrac{\mathbb{E}_2 \cosh^\theta \tilde g(\beta', J, \tilde q_1, \tilde q_0) \tanh \tilde g(\beta', J,\tilde q_1, \tilde q_0)}{\mathbb{E}_2 \cosh^\theta \tilde g(\beta', J,\tilde q_1, \tilde q_0)}\right]^2.
\end{array}
\label{eq:self_RSB}
\end{equation}

On the other hand, in 2RSB assumption the order parameters follows \eqref{2RSBass}. The quenched free energy in 2RSB assumption for P-spin model is 
\begin{align}
     -\beta f_{2RSB}(\q_2, \q_1, \q_0 \vert \beta, J, \theta_2, \theta_1) =& \ln 2 + \dfrac{{\beta'}^2 J^2 (P-1)}{4} \left[ 1-(\theta_2-1) \q_2^{P} + (\theta_2-\theta_1) \q_1^P + \theta_1 \q_0^P \right] \notag \\
     & - \dfrac{{\beta'}^2 J^2 P}{2} \q_2^{P-1}+ \dfrac{1}{\theta_1} \mathbb{E}_1 \log \mathbb{E}_2 \left[ \mathbb{E}_3 \cosh^{\theta_2} g(\beta, J, \q_2, \q_1, \q_0) \right]^{\theta_1/\theta_2}
\end{align}
where 
\begin{align}
     g(\beta', J, \q_1, \q_0) =\b J z^{(1)} \sqrt{\dfrac{P}{2} \q_0^{P-1}} + \b J z^{(2)} \sqrt{\dfrac{P}{2} (\q_1^{P-1}-\q_0^{P-1})}+ \b J z^{(3)} \sqrt{\dfrac{P}{2} (\q_2^{P-1}-\q_1^{P-1})},
\end{align}
and $\mathbb{E}_a$ is the average w.r.t. $z^{(a)}$, for $a=1,2,3$.

From now on, we imply the dependence of $g$, $f_{1RSB}$ and $f_{2RSB}$ on $\beta$ and $J$. 
The order parameters $\q_0, \ \q_1, \ \q_2$ fulfill the following self-consistency equations 
\begin{align}
    \q_0=& \mathbb{E}_1 \left\{ \dfrac{\mathbb{E}_2 \left[\left( \mathbb{E}_3 \cosh^{\theta_2} g(\q_2,\q_1, \q_0)\right)^{\theta_1/\theta_2} \dfrac{\mathbb{E}_3 \cosh^{\theta_2}g(\q_2, \q_1, \q_0) \tanh g(\q_2, \q_1, \q_0) }{\mathbb{E}_3 \cosh^{\theta_2}g(\q_2, \q_1, \q_0)}\right]}{\mathbb{E}_2 \left[ \mathbb{E}_3 \cosh^{\theta_2} g(\q_2,\q_1, \q_0)\right]^{\theta_1/\theta_2}}\right\}^2 \label{eq:q0_PSK}\\
    \q_1=& \mathbb{E}_1 \left\{ \dfrac{\mathbb{E}_2 \left[\left( \mathbb{E}_3 \cosh^{\theta_2} g(\q_2,\q_1, \q_0)\right)^{\theta_1/\theta_2} \dfrac{\mathbb{E}_3 \cosh^{\theta_2}g(\q_2, \q_1, \q_0) \tanh g(\q_2, \q_1, \q_0) }{\mathbb{E}_3 \cosh^{\theta_2}g(\q_2, \q_1, \q_0)}\right]^2}{\mathbb{E}_2 \left[ \mathbb{E}_3 \cosh^{\theta_2} g(\q_2,\q_1, \q_0)\right]^{\theta_1/\theta_2}}\right\} \\
    \q_2=& \mathbb{E}_1 \left\{ \dfrac{\mathbb{E}_2 \left[\left( \mathbb{E}_3 \cosh^{\theta_2} g(\q_2,\q_1, \q_0)\right)^{\theta_1/\theta_2} \dfrac{\mathbb{E}_3 \cosh^{\theta_2}g(\q_2, \q_1, \q_0) \tanh^2 g(\q_2, \q_1, \q_0) }{\mathbb{E}_3 \cosh^{\theta_2}g(\q_2, \q_1, \q_0)}\right]}{\mathbb{E}_2 \left[ \mathbb{E}_3 \cosh^{\theta_2} g(\q_2,\q_1, \q_0)\right]^{\theta_1/\theta_2}}\right\}.
    \label{eq:q2_PSK}
\end{align}

We decide, also in this case, to develop the computations for $\theta_2 \to 1$. The same considerations apply here as in SK model. So, our scope is to prove that for values of $\theta_2$ close but away from one, the 2RSB expression of the quenched 
free energy is 
smaller
than the 1RSB expression, i.e. 
$f_{2RSB}(\bar{q}_2, \bar{q}_{1},\bar{q}_0|\theta)<f_{1RSB}(\tilde{q}_1, \tilde q_0)$, 
below a critical line in the parameters space $(J,\beta)$.  Therefore, we want to exploit the 2RSB quenched free-energy as in \eqref{eq:expansionSK_t1} and, to do so, we need to expand near $\theta_2=1$, the expressions of the self consistence equations \eqref{eq:q0_PSK}-\eqref{eq:q2_PSK}, as in \eqref{q0_expSK}, \eqref{q1_expSK}, \eqref{q2_expSK}, respectively (the values of $A(\q_0, \q_1, \q_2), B(\q_0, \q_1, \q_2), C(\q_0, \q_1, \q_2)$ have the same functional expression of SK ones found in Appendix \ref{app:correction}). 

Computing the derivative we need in the expansion \eqref{eq:expansionSK_t1} for the P-spin model we get 
\begin{align}
    K(\tilde q_0, \tilde q_1, \tilde q_2(\tilde q_1, \tilde q_0)) =& \dfrac{{\beta '}^2 J^2}{4} (P-1) \left( \tilde q_1^P - \tilde q_2^P \right) - \mathbb{E}_1 \dfrac{\mathbb{E}_2 \left[ \left( \mathbb{E}_3 \cosh g(\tilde q_0, \tilde q_1, \tilde q_2) \right)^{\theta_1} \ln \mathbb{E}_3 \cosh g(\tilde q_0, \tilde q_1, \tilde q_2) \right]}{\mathbb{E}_2 \left[ \mathbb{E}_3 \cosh g(\tilde q_0, \tilde q_1, \tilde q_2)\right]^{\theta_1}} \notag \\
    &+ \mathbb{E}_1 \dfrac{\mathbb{E}_2 \left[ \left( \mathbb{E}_3 \cosh g(\tilde q_0, \tilde q_1, \tilde q_2) \right)^{\theta_1}\dfrac{\mathbb{E}_3 \cosh g(\tilde q_0, \tilde q_1, \tilde q_2) \ln \cosh g(\tilde q_0, \tilde q_1, \tilde q_2) }{\mathbb{E}_3 \cosh g(\tilde q_0, \tilde q_1, \tilde q_2)}\right]}{\mathbb{E}_2 \left[ \mathbb{E}_3 \cosh g(\tilde q_0, \tilde q_1, \tilde q_2)\right]^{\theta_1}}. 
    \label{K_PSK}
\end{align}

Then, following the same route of SK model, we suppose that the extremum is global and we find 
\begin{align}
    \partial_x  K(\tilde q_0, \tilde q_1, x) = - \dfrac{{\beta '}^2 J^2}{4} (P-1) P&  (\tilde q_2 (\tilde q_1, \tilde q_0))^{P-1} \notag \\
    &\cdot\left\{ x- \mathbb{E}_1 \left[\dfrac{\mathbb{E}_2 \cosh^{\theta_1} \tilde g(\tilde q_1, \tilde q_0)\dfrac{\mathbb{E}_3 \cosh g (\q_0, \q_1, x ) \tanh^2 g (\q_0, \q_1, x)}{\mathbb{E}_3 \cosh^{\theta_1} g (\q_0, \q_1, \q_2)}} {\mathbb{E}_2 \cosh^{\theta_1} \tilde g(\q_1, \q_0)}\right]\right\}
\end{align}
which is nullified when $x$ is exactly $\tilde q_2(\tilde q_0, \tilde q_1)$. 

\begin{align}
    \partial_{x^2}  &K(\tilde q_0, \tilde q_1, x) \Big\vert_{x= \tilde q_2(\tilde q_0, \tilde q_1)} = - \dfrac{{\beta '}^2 J^2}{4} P(P-1) (\tilde q_2(\tilde q_0, \tilde q_1))^{P-2} \notag \\
    &\cdot\left\{ 1-{\beta '}^2 J^2 \dfrac{P(P-1)}{2} (\tilde q_2(\tilde q_0, \tilde q_1))^{P-2} \mathbb{E}_1 \left[ \dfrac{\mathbb{E}_2 \cosh^{\theta_1}  \tilde g(\tilde q_1, \tilde q_0) \dfrac{\mathbb{E}_3 \sech^3 g (\q_0, \q_1, \tilde q_2(\tilde q_0, \tilde q_1) ) }{\mathbb{E}_3 \cosh g (\q_0, \q_1, \tilde q_2(\tilde q_0, \tilde q_1) ) }}{\mathbb{E}_2 \left( \cosh^{\theta_1}  \tilde g(\tilde q_1, \tilde q_0)\right)}\right]\right\}, 
\end{align}

which corresponds to assume that the critical line, when $\tilde q_2(\tilde q_1, \tilde q_0) \to \tilde q_1$, is 

\begin{align}
        1-{\beta '}^2 J^2  \dfrac{P(P-1)}{2} \tilde q_1^{P-2}\mathbb{E}_1 \left[ \dfrac{\mathbb{E}_2 \cosh^{\theta_1-4}  \tilde g(\tilde q_1, \tilde q_0)}{\mathbb{E}_2 \cosh^{\theta_1}  \tilde g(\tilde q_1, \tilde q_0)}\right] <0.
\end{align}

Moreover, if we suppose that $\tilde q_0$ is null, we reach the critical line of Gardner temperature \cite{gardner1985spin}: 
\begin{align}
            1-\beta^2 J^2\dfrac{P(P-1)}{2} \tilde q_1^{P-2}\dfrac{\mathbb{E}_2 \cosh^{\theta_1-4} \left( \beta ' \sqrt{\dfrac{P}{2} \tilde q_1^{P-1}} z^{(2)}\right)}{\mathbb{E}_2 \cosh^{\theta_1} \left( \beta ' \sqrt{\dfrac{P}{2} \tilde q_1^{P-1}} z^{(2)}\right)} <0.
\end{align}

In $\theta_1 \to 0$ and $\theta_2 \to \theta_1$ limits instead we get respectevely
\begin{align}
    1-(\beta ')^2 J^2 &\dfrac{P(P-1)}{2} (\tilde q_2(\tilde q_0, \tilde q_1))^{P-2} \mathbb{E}_1 \left\{ \mathbb{E}_2 \left[ \dfrac{\mathbb{E}_3 \cosh^{\theta_2}\tilde g (\tilde q_1, \tilde q_0, \tilde q_2(\tilde q_0, \tilde q_1) )\textrm{sech}^2 g(\tilde q_1, \tilde q_0, \tilde q_2(\tilde q_0, \tilde q_1) ) }{\mathbb{E}_3 \cosh^{\theta_2} g(\tilde q_1, \tilde q_0, \tilde q_2(\tilde q_0, \tilde q_1) )}\right]\right. \notag \\
    &+ \theta_2 \left[  \mathbb{E}_2 \left( \dfrac{\mathbb{E}_3 \cosh^{\theta_2} g(\tilde q_1, \tilde q_0, \tilde q_2(\tilde q_0, \tilde q_1) ) \tanh^2 g(\tilde q_1, \tilde q_0, \tilde q_2(\tilde q_0, \tilde q_1) )}{\mathbb{E}_3 \cosh^{\theta_2} g(\tilde q_1, \tilde q_0, \tilde q_2(\tilde q_0, \tilde q_1) ) } \right) \right. \notag \\
    &\left. \left.- \mathbb{E}_2 \left( \dfrac{\mathbb{E}_3 \cosh^{\theta_2} g(\tilde q_1, \tilde q_0, \tilde q_2(\tilde q_0, \tilde q_1) ) \tanh g(\tilde q_1, \tilde q_0, \tilde q_2(\tilde q_0, \tilde q_1) )}{\mathbb{E}_3 \cosh^{\theta_2} g(\tilde q_1, \tilde q_0, \tilde q_2(\tilde q_0, \tilde q_1) ) } \right)^2 \right]\right\}^2 <0,
\end{align}
and 
\begin{align}
1-(\beta ')^2 J^2 &\dfrac{P(P-1)}{2} (\tilde q_2)^{P-2} \mathbb{E}_1 \left\{ \dfrac{1}{\mathbb{E}_2 \mathbb{E}_3 \cosh^{\theta_1} g(\tilde q_1, \tilde q_0, \tilde q_2(\tilde q_0, \tilde q_1) )}\mathbb{E}_2 \left\{ \dfrac{1}{\mathbb{E}_3 \cosh^{\theta_1} g(\tilde q_1, \tilde q_0, \tilde q_2(\tilde q_0, \tilde q_1) )}\right.\right.\notag \\
    &\cdot \left[ \mathbb{E}_3 \cosh^{\theta_1}g(\tilde q_1, \tilde q_0, \tilde q_2(\tilde q_0, \tilde q_1) )\sech^2 g(\tilde q_1, \tilde q_0, \tilde q_2(\tilde q_0, \tilde q_1) ) \right. \notag \\
    &+ \theta_1 \left( \mathbb{E}_3 \cosh^{\theta_1} g(\tilde q_1, \tilde q_0, \tilde q_2(\tilde q_0, \tilde q_1) ) \tanh^2 g(\tilde q_1, \tilde q_0, \tilde q_2(\tilde q_0, \tilde q_1) ) \right.\notag \\
&\left.\left.\left.\left.- \dfrac{\left( \mathbb{E}_3 \cosh^{\theta_1} g(\tilde q_1, \tilde q_0, \tilde q_2(\tilde q_0, \tilde q_1) ) \tanh g(\tilde q_1, \tilde q_0, \tilde q_2(\tilde q_0, \tilde q_1) )\right)^2}{\mathbb{E}_3 \cosh^{\theta_1}g(\tilde q_1, \tilde q_0, \tilde q_2(\tilde q_0, \tilde q_1) )}\right) \right]^2\right\}\right\} <0,
\end{align}
with the same considerations developped in SK model. 

\section{Generalizing to $K$RSB with $K$ finite}
\label{sec:krsb}
Now we wonder if the procedure shown in the previous Section could be apply also for $K$th step of RSB, with $K$ finite. We will see all the computations for SK model and we will give only the results for P-spin ones. 

In $K$RSB case the probability distribution for the order parameter, the overlap $q$ can be written as 
\begin{align}
    \lim_{N \to +\infty} P'_N(q)= \sum_{a=0}^K \left( \theta_{a+1} - \theta_{a} \right)\delta \left( q-\q_{a}\right)
\end{align}
supposing $\theta_0=0$ and $\theta_{K+1}=1$. 
As already shown in \cite{AABO-JPA2020}, the quenched statistical pressure of the SK model is 
\begin{align}
    \mathcal A^{KRSB}= \log 2 + \dfrac{1}{\theta_1} \mathbb{E}_1 \log \mathcal N_1 - \dfrac{\beta^2 J^2}{4} \left[ \sum_{a=0}^K \left( \theta_{a+1}-\theta_a \right) \q_{a}^2 -1\right] - \dfrac{\beta^2 J^2}{2} \left( 1- \q_{K}\right)
\end{align}
with 
\begin{align}
    \mathcal{N}_a = \begin{cases}
        \mathbb{E}_{a+1} \left( \mathcal{N}_{a+1} \right)^{\theta_a/\theta_{a+1}} \ \ \ a=1, \hdots , K\\
         \cosh \left( \beta \sum_{a=0}^K z^{(a+1)} 
 \sqrt{\q_a-\q_{a-1}}\right), \ \ \ a=K+1,
    \end{cases}
\end{align}
where $\q_{a-1}=0$. 
The order parameters fulfill the following self-consistency equations: 

\begin{align}
\q_0 &=  \mathbb{E}_1 \left[\frac{1}{\mathcal{N}_1}\mathbb{E}_2 \left[ \mathcal{N}_2^{\frac{\theta_1}{\theta_2}-1} \cdots \mathbb{E}_{K+1}\left[ \cosh^{\theta_K} g(\bm z) \tanh g(\bm z)  \right] \right]\right]^2 \\
\q_1 &=  \mathbb{E}_1 \left[\frac{1}{\mathcal{N}_1}\left[\mathbb{E}_2 \left[ \mathcal{N}_2^{\frac{\theta_1}{\theta_2}-1} \cdots \mathbb{E}_{K+1}\left[ \cosh^{\theta_K} g(\bm z) \tanh g(\bm z)  \right] \right]\right]^2\right] \\
... \notag \\
\q_{K} &=  \mathbb{E}_1 \left[\frac{1}{\mathcal{N}_1}\mathbb{E}_2 \left[ \mathcal{N}_2^{\frac{\theta_1}{\theta_2}-1} \cdots \mathbb{E}_{K+1}\left[ \cosh^{\theta_K} g(\bm z) \tanh^2 g(\bm z)  \right] \right]\right].
\end{align}
where $g(\bm z) = \left( \beta J \sum_{a=0}^K z^{(a+1)} \sqrt{\q_a-\q_{a-1}}\right)$.

Instead, in the $(K-1)$RSB case the probability distribution is 
\begin{align}
    \lim_{N \to +\infty} P'_N(q)= \sum_{a=0}^{K-1} \left( \theta_{a+1} - \theta_{a} \right)\delta \left( q-\q_{a}\right)
\end{align}
supposing $\theta_0=0$ and $\theta_{K}=1$. The quenched statistical pressure of the SK model in (K-1)RSB assumption is 
\begin{align}
    \mathcal A^{(K-1)RSB}= \log 2 + \dfrac{1}{\theta_1} \mathbb{E}_1 \log \tilde{\mathcal N}_1 - \dfrac{\beta^2 J^2}{4} \left[ \sum_{a=0}^{K-1} \left( \theta_{a+1}-\theta_a \right) \tilde q_{a}^2 -1\right] - \dfrac{\beta^2 J^2}{2} \left( 1- \tilde q_{K-1}\right)
\end{align}

with 
\begin{align}
    \tilde{\mathcal{N}}_a = \begin{cases}
        \mathbb{E}_{a+1} \left( \tilde{\mathcal{N}}_{a+1} \right)^{\theta_a/\theta_{a+1}} \ \ \ a=1, \hdots , K-1\\
         \cosh \left( \beta \sum_{a=0}^{K-1} z^{(a+1)} 
 \sqrt{\tilde q_a-\tilde q_{a-1}}\right), \ \ \ a=K.
    \end{cases}
\end{align}
The order parameter fulfills the following self-consistency equations: 

\begin{align}
\tilde q_0 &=  \mathbb{E}_1 \left[\frac{1}{\tilde{\mathcal{N}}_1}\mathbb{E}_2 \left[ \tilde{\mathcal{N}}_2^{\frac{\theta_1}{\theta_2}-1} \cdots \mathbb{E}_{K}\left[ \cosh^{\theta_{K-1}} \tilde g(\bm z) \tanh \tilde g(\bm z)  \right] \right]\right]^2 \\
\tilde q_1 &=  \mathbb{E}_1 \left[\frac{1}{\tilde{\mathcal{N}}_1}\left[\mathbb{E}_2 \left[ \tilde{\mathcal{N}}_2^{\frac{\theta_1}{\theta_2}-1} \cdots \mathbb{E}_{K}\left[ \cosh^{\theta_{K-1}} \tilde g(\bm z) \tanh \tilde g(\bm z)  \right] \right]\right]^2\right] \\
... \notag \\
\tilde q_{K-1} &=  \mathbb{E}_1 \left[\frac{1}{\tilde{\mathcal{N}}_1}\mathbb{E}_2 \left[ \tilde{\mathcal{N}}_2^{\frac{\theta_1}{\theta_2}-1} \cdots \mathbb{E}_{K}\left[ \cosh^{\theta_{K-1}} \tilde g(\bm z) \tanh^2 \tilde g(\bm z)  \right] \right]\right] 
\end{align}
where $\tilde g(\bm z)=\left( \beta J \sum_{a=0}^{K-1} z^{(a+1)} 
 \sqrt{\q_a-\q_{a-1}}\right)$.

As we have already seen in the $2$RSB case, we come back to $(K-1)$RSB approximation for $\theta_K \to 1$, $\theta_a \to \theta_{a-1}, \ a=2, \hdots, K $, and $\theta_1 \to 0$. We can apply the same procedure for each case: we will see further only the former. \\
 Our purpose now is to prove that, for small but finite values close to $\theta_K=1$, we have a particular condition such that $f_{KRSB}<f_{(K-1)RSB}$: 
 \begin{align}
     f_{KRSB}(\bm q \vert \beta, J, \bm \theta)= f_{(K-1)RSB} (\bm {\tilde q} \vert \beta, J, \bm {\tilde \theta}) + (\theta_K -1) \dfrac{\partial f_{KRSB}(\bm q \vert \beta, J, \bm \theta)}{\partial \theta_K} \vert_{\theta_K=1}<  f_{(K-1)RSB} (\bm {\tilde q} \vert \beta, J, \bm {\tilde \theta}).
 \end{align}

 First of all, we need to expand the self-consistency equation of $\q_1, \hdots \q_{K}$ as
 \begin{align}
     \q_a &= \tilde q_a  + (\theta_K-1) A_a(\bm{\tilde q} \vert \beta, J, \bm{\tilde J}) \ \ \ a=0, \hdots, K
 \end{align}
where $\tilde q_a, \ a=0, \hdots, K-1$ fulfill the self-consistency equations of $(K-1)$RSB approximation, whereas $\tilde q_K$ is the solution of the following self-consistency equation
\begin{align}
    \tilde q_K = \mathbb{E}_1 \left[ \dfrac{1}{\tilde{\mathcal N}_1 } \mathbb{E}_2 \left[ \tilde{\mathcal{N}}_2^{\frac{\theta_1}{\theta_2}-1} \hdots \mathbb{E}_{K+1} \cosh g(\bm z) \tanh^2 g(\bm z) \right]\right].
\end{align}
The correction terms $A_a(\bm{\tilde q} \vert \beta, J, \tilde J)$, $a=1, \hdots, K$, can be computed using the steps described in Appendix \ref{app:correction}.

Computing the derivative of $\mathcal A_{KRSB}$ w.r.t. $\theta_K$, where $\theta_K=1$, we get
\begin{align}
    K(\bm{\tilde q}) =& \dfrac{\partial \mathcal A_{KRSB}}{\partial \theta_K} \Big\vert_{\theta_K=1} = \dfrac{\beta^2 J^2}{4} \left( \tilde q_{K-1}^2-\tilde q_{K}\right) \notag \\
    &+ \mathbb{E}_1 \left[ \dfrac{1}{\mathcal N_1} \mathbb{E}_2 \mathcal{N}_2^{\frac{\theta_1}{\theta_2}-1} \hdots \mathbb{E}_K \left( \left( \mathbb{E}_{K+1} \cosh g(\bm z) \right)^{\theta_{K-1}-1}\mathbb{E}_{K+1} \cosh g(\bm z) \log \cosh g(\bm z) \right)\right] \notag \\
    &- \mathbb{E}_1 \left[ \dfrac{1}{\mathcal{N}_1} \mathbb{E}_2 \mathcal{N}_2^{\frac{\theta_2}{\theta_1}-1} \hdots \mathbb{E}_K \left( \left( \mathbb{E}_{K+1} \cosh g(\bm z) \right)^{\theta_{K-1}}\log \mathbb{E}_{K+1} \cosh g(\bm z) \right)\right]
\end{align}

which is equal to zero when $\tilde q_K = \tilde q_{K-1}$.
To inspect the sign of $K(\bm{\tilde q})$, we are going to study the behaviour of the function $ K(\bm{\tilde q}_{a=0, \hdots, K-1}, x)$ for $x \in [0, \tilde q_{K-1}]$. We notice that for $x=\tilde q_{K-1}$ we get $K(\bm{\tilde q}_{a=0, \hdots, K-1}, \tilde q_{K-1})=0$. The extremum of the function, supposing it is the global one, is 

\begin{align}
    \partial_x K(\bm{\tilde q}_{a=0, \hdots, K-1}, x) = - \dfrac{\beta^2 J^2}{2} \left( x - \mathbb{E}_1 \left[ \dfrac{1}{\tilde{\mathcal N}_1 } \mathbb{E}_2 \left[ \tilde{\mathcal{N}}_2^{\frac{\theta_1}{\theta_2}-1} \hdots \mathbb{E}_{K+1} \cosh g(\bm z) \tanh^2 g(\bm z) \right]\right]\right)
\end{align}
which is null when $x$ assumes the value of $\tilde q_K(\bm \tilde{q})$.

\begin{align}
    \partial_{x^2} &K(\bm{\tilde q}_{a=0, \hdots, K-1}, x) \vert_{x= \tilde q_K} = -\dfrac{\beta^2 J^2}{2} \notag \\
    &\cdot\left( 1- \beta^2 J^2\mathbb{E}_1 \left[ \dfrac{1}{\mathcal N_1 } \mathbb{E}_2 \mathcal{N}_2^{\frac{\theta_1}{\theta_2}-1} \hdots \mathbb{E}_K \left( \left( \mathbb{E}_{K+1} \cosh g(\bm z) \right)^{\theta_{K-1}} \mathbb{E}_{K+1} \sech^3 g(\bm z) \right)\right]\right).
\end{align}
If we put $\tilde q_K \to \tilde q_{K-1}$ we recover the corresponding of AT line for $K$ RSB, with $K$ finite: 
\begin{align}
    1-\beta^2 J^2 \mathbb{E}_1 \left[ \dfrac{1}{\mathcal N_1} \mathbb{E}_2 \mathcal{N}_2^{\frac{\theta_1}{\theta_2}-1} \hdots \mathbb{E}_K \cosh^{\theta_{K-1}-4} \tilde g(\bm z)\right]<0.
\end{align}

\par\medskip
In the case of Ising $P$ spin model, the computations are analogous and the corresponding AT line for $K$RSB is 
\begin{align}
    1-{\b}^2 \dfrac{P(P-1)}{2} \left( \tilde q_K\right)^{P-2} \mathbb{E}_1 \left[ \dfrac{1}{\mathcal N_1} \mathbb{E}_2 \mathcal{N}_2^{\frac{\theta_1}{\theta_2}-1} \hdots \mathbb{E}_K \cosh^{\theta_{K-1}-4} \tilde g(\bm z)\right]<0.
\end{align}

\section{Conclusions and outlooks}

In this Paper we have devise with a new method, inspired to the work by Toninelli in \cite{toninelli2002almeida}, the expressions of the critical line between 2RSB and 1RSB approximations for SK model and P-spin model. In particular, we have seen all the computations for the $\theta_2 \to 1$ limit and we have shown only the results for $\theta_1 \to 0$ and $\theta_2 \to \theta_1$. 
In these different limits we have found different analytical expressions which collapse only in RS case, in the well-known AT line. Indeed, it is possible to show coherence with the results in the 1RSB vs RS case in \cite{albanese2023almeida}. All these results represent a novelty in the Literature where the most used technique is that one inherited by the work by de Almeida and Thouless inspired by replica approach (apart from the work by Gardner and Chen \cite{gardner1985spin, chen2017variational}. The method applied in this work, instead, seems to be rigorous and straightforward, not based on replica approach, and suitable for other typologies of models, also not part of the spin glasses.\\ 
One of the possible future outlook of this work, indeed, is to extend the technique also for associative neural networks, as already done for the classic AT line in \cite{albanese2023almeida}. 
The final purpose will be to better understand the transition between the different steps of RSB in different type of models in order to deepen the phenomenology under that. 

\acknowledgments

The author gratefully acknowledges useful conversations with Adriano Barra and Andrea Alessandrelli. \\
The author acknowledges the PRIN grant {\em Statistical Mechanics of Learning Machines} n. 20229T9EAT for financial support.\\
The author is member of the group GNFM of INdAM which is acknowledged too. 


\begin{thebibliography}{10}

\bibitem{AABO-JPA2020}
E.~Agliari, L.~Albanese, A.~Barra, and G.~Ottaviani.
\newblock Replica symmetry breaking in neural networks: A few steps toward
  rigorous results.
\newblock {\em Journal of Physics A: Mathematical and Theoretical}, 53, 2020.

\bibitem{GuysAlone}
L.~Albanese and A.~Alessandrelli.
\newblock On gaussian spin glass with p-wise interactions.
\newblock {\em Journal of Mathematical Physics}, 63:43302, 2022.

\bibitem{albanese2023almeida}
L.~Albanese, A.~Alessandrelli, A.~Annibale, and A.~Barra.
\newblock About the de {A}lmeida--{T}houless line in neural networks.
\newblock {\em Physica A: Statistical Mechanics and its Applications},
  633:129372, 2024.

\bibitem{bardina2004}
X.~Bardina, D.~M\'arquez-Carreras, C.~Rovira, and S.~Tindel.
\newblock The p-spin interaction model with external field.
\newblock {\em Potential Analysis}, 21:311–362, 2004.

\bibitem{chen2017variational}
W.-K. Chen.
\newblock Variational representations for the parisi functional and the
  two-dimensional guerra--talagrand bound.
\newblock {\em Annals of Probability}, 45, 2017.

\bibitem{chen2021almeida}
W.-K. Chen.
\newblock On the {A}lmeida-{T}houless transition line in the
  {S}herrington-{K}irkpatrick model with centered gaussian external field.
\newblock {\em Electronic Communications in Probability}, 26:1--9, 2021.

\bibitem{coolen2001statistical}
A.~Coolen.
\newblock Statistical mechanics of recurrent neural networks i—statics.
\newblock In {\em Handbook of biological physics}, volume~4, pages 553--618.
  Elsevier, 2001.

\bibitem{Crisanti-RSB}
A.~Crisanti, D.~J. Amit, and H.~Gutfreund.
\newblock Saturation level of the {H}opfield model for {N}eural {N}etwork.
\newblock {\em Europhysics Letters (EPL)}, 2:337--341, 8 1986.

\bibitem{de1978stability}
J.~R. de~Almeida and D.~J. Thouless.
\newblock Stability of the {S}herrington-{K}irkpatrick solution of a spin glass
  model.
\newblock {\em Journal of Physics A: Mathematical and General}, 11(5):983,
  1978.

\bibitem{gardner1985spin}
E.~Gardner.
\newblock Spin glasses with p-spin interactions.
\newblock {\em Nuclear Physics B}, 257:747--765, 1985.

\bibitem{ghirlanda1998general}
S.~Ghirlanda and F.~Guerra.
\newblock General properties of overlap probability distributions in disordered
  spin systems. towards parisi ultrametricity.
\newblock {\em Journal of Physics A: Mathematical and General}, 31(46):9149,
  1998.

\bibitem{gradenigo2020solving}
G.~Gradenigo, M.~C. Angelini, L.~Leuzzi, and F.~Ricci-Tersenghi.
\newblock Solving the spherical p-spin model with the cavity method:
  equivalence with the replica results.
\newblock {\em Journal of Statistical Mechanics: Theory and Experiment},
  2020(11):113302, 2020.

\bibitem{guerra_broken}
F.~Guerra.
\newblock Broken replica symmetry bounds in the mean field spin glass model.
\newblock {\em Communications in Mathematical Physics}, 233:1--12, 2003.

\bibitem{guerra2006replica}
F.~Guerra.
\newblock The replica symmetric region in the {S}herrington-{K}irkpatrick mean
  field spin glass model. the {A}lmeida-{T}houless line.
\newblock {\em arXiv preprint cond-mat/0604674}, 2006.

\bibitem{guerra2002thermodynamic}
F.~Guerra and F.~L. Toninelli.
\newblock The thermodynamic limit in mean field spin glass models.
\newblock {\em Communications in Mathematical Physics}, 230:71--79, 2002.

\bibitem{Gavin-PRE2022}
G.~S. Hartnett, E.~Parker, and E.~Geist.
\newblock Replica symmetry breaking in bipartite spin glasses and neural
  networks.
\newblock {\em Physical Review E}, 98(2):022116, 2018.

\bibitem{imparato2006evaluation}
A.~Imparato and L.~Peliti.
\newblock Evaluation of free energy landscapes from manipulation experiments.
\newblock {\em Journal of Statistical Mechanics: Theory and Experiment},
  2006(03):P03005, 2006.

\bibitem{MPV}
M.~Mézard, G.~Parisi, and M.~A. Virasoro.
\newblock {\em Spin glass theory and beyond: An Introduction to the Replica
  Method and Its Applications}, volume~9.
\newblock World Scientific Publishing Company, 1987.

\bibitem{nishimori2001statistical}
H.~Nishimori.
\newblock {\em Statistical physics of spin glasses and information processing:
  an introduction}.
\newblock Number 111. Clarendon Press, 2001.

\bibitem{panchenko2013parisi}
D.~Panchenko.
\newblock The {P}arisi ultrametricity conjecture.
\newblock {\em Annals of Mathematics}, pages 383--393, 2013.

\bibitem{peliti1994collective}
L.~Peliti and U.~Bastolla.
\newblock Collective adaptation in a statistical model of an evolving
  population.
\newblock {\em Comptes rendus de l'Academie des sciences. Serie III, Sciences
  de la vie}, 317(4):371--374, 1994.

\bibitem{sherrington1975solvable}
D.~Sherrington and S.~Kirkpatrick.
\newblock Solvable model of a spin-glass.
\newblock {\em Physical review letters}, 35(26):1792, 1975.

\bibitem{Steffan-RSB}
H.~Steffan and R.~Kühn.
\newblock Replica symmetry breaking in attractor neural network models.
\newblock {\em Zeitschrift für Physik B Condensed Matter}, 95, 1994.

\bibitem{talagrand2006parisi}
M.~Talagrand.
\newblock The parisi formula.
\newblock {\em Annals of mathematics}, pages 221--263, 2006.

\bibitem{talagrand2010construction}
M.~Talagrand.
\newblock Construction of pure states in mean field models for spin glasses.
\newblock {\em Probability theory and related fields}, 148:601--643, 2010.

\bibitem{toninelli2002almeida}
F.~L. Toninelli.
\newblock About the {A}lmeida-{T}houless transition line in the
  {S}herrington-{K}irkpatrick mean-field spin glass model.
\newblock {\em EPL (Europhysics Letters)}, 60(5):764, 2002.

\end{thebibliography}

\appendix
\section{Correction terms}
\label{app:correction}

In this Section we will sketch how to find the correction terms in a general finite step of Replica Symmetry Breaking. Afterwards, we will apply the technique in the case of $2$RSB to find $A(\q_0, \q_1, \q_2)$, $B(\q_0, \q_1, \q_2)$, $C(\q_0, \q_1, \q_2)$ for Section \ref{sec:skt2}. 

Let us consider the $K$RSB step. We can express the self-consistency equations as 
\begin{align}
    \begin{cases}
        \q_B=& \mathbb{E}_1 \left[ \dfrac{1}{\mathcal N_1} \mathbb{E}_2 \left[\mathcal{N}_2^{\theta_2/\theta_1 -1} \hdots \right. \right . \\
        &\left. \left.\hspace{1cm}\cdot\mathbb{E}_{B+1} \mathcal{N}_{B+1}^{\theta_{B+1}/\theta_B } \left[\dfrac{\mathbb{E}_{B+2} \mathcal{N}_{B+2}^{\theta_{B+2}/\theta_{B+1} -1} \hdots \mathbb{E}_{K+1} \cosh^{\theta_K} g(\bm z) \tanh g(\bm z) \hdots}{\mathcal{N}_{B+1}} \right]^2 \hdots\right] \right]\\
        \q_K=&  \mathbb{E}_1 \left[\dfrac{1}{\mathcal{N}_1}\mathbb{E}_2 \left[ \mathcal{N}_2^{\frac{\theta_1}{\theta_2}-1} \cdots \mathbb{E}_{K+1}\left[ \cosh^{\theta_K} g(\bm z) \tanh^2 g(\bm z)  \right] \right]\right], 
    \end{cases}
    \label{eq:selfAPP}
\end{align}
with $B=0, \hdots, K-1$. 
In order to do the expansion w.r.t. $\theta_K$, we need to highlight that every $\mathcal N_a$ depends on $\theta_K$ (since they depend on $\mathcal{N}_K$ in iterative way). Moreover, we notice that 
\begin{align}
\label{eq:Na1}
\mathcal N_{a-1}= \mathbb{E}_a \mathcal{N}_a^{\frac{\theta_{a-1}}{\theta_a}},
\end{align} $a=2,\hdots, K+1$,
so the expansion of $\mathcal N_{a-1}$ and $\mathcal{N}_a^{\frac{\theta_{a-1}}{\theta_a}}$ w.r.t. $\theta_K$ around $\theta_K=1$ will be the same (apart from the average w.r.t. $J^{(a)}$). 
Then, we need to recall that, for any function $\alpha, \beta, \gamma, \delta$ independent from $\theta_K$ we get at first order around $\theta_K=1$
\begin{align}
    (\alpha+\beta(\theta_K-1))(\gamma+\delta(\theta_K-1))=& \alpha \gamma + (\theta_K-1) (\beta \gamma+\alpha \delta), \label{eq:expansionmult} \\
    \dfrac{ (\alpha+\beta(\theta_K-1))}{(\gamma+\delta(\theta_K-1))} =& \dfrac{\alpha}{\gamma} + (\theta_K-1) \dfrac{\beta \gamma-\alpha\delta}{\gamma^2}, \label{eq:expansiondiv}\\
    (\alpha+\beta(\theta_K-1))^2=& \alpha^2 + 2\alpha \beta(\theta_K-1).\label{eq:expansionsquare}
\end{align}

From now on, we now consider only the truncation of the expansion at the first order. 
For each of the self-consistency equation in \eqref{eq:selfAPP} we need to apply an iterative procedure which involves the application of at least one of \eqref{eq:expansionmult}-\eqref{eq:expansionsquare} at each step. 
\begin{enumerate}
    \item Let $B \in \{0, 1, \hdots, K\}$ fixed. As one can see from the expressions, the position of the square depends on $B$. If $B=K$, we can start from the following expansion 
    \begin{align}
        &\dfrac{\mathbb{E}_{K+1} \cosh^{\theta_K} g(\bm z) \tanh^2 g(\bm z)}{\mathbb{E}_{K+1} \cosh^{\theta_K} g(\bm z)} \sim \dfrac{\mathbb{E}_{K+1} \sinh g(\bm z) \tanh^2 g(\bm z)}{\mathbb{E}_{K+1}\cosh g(\bm z)} + (\theta_K-1) \notag \\
    &\hspace{2cm}\cdot \left[ \dfrac{\mathbb{E}_{K+1} \log \cosh (\bm z)\sinh g(\bm z) \tanh g(\bm z) \mathbb{E}_{K+1} \cosh g(\bm z) \log \cosh (\bm z)}{\left( \mathbb{E}_{K+1}\cosh g(\bm z)\right)^2}\right] \label{eq:SCEsquare}.
    \end{align}
    If $B \neq K$, there is no square on the hyperbolic tangent so we start from 
    \begin{align}
        &\dfrac{\mathbb{E}_{K+1} \cosh^{\theta_K} g(\bm z) \tanh g(\bm z)}{\mathbb{E}_{K+1} \cosh^{\theta_K} g(\bm z)} \sim \tanh \tilde g(\bm z) + (\theta_K-1) \notag \\
    &\hspace{2cm}\cdot\left[ \dfrac{\mathbb{E}_{K+1} \sinh g(\bm z) \log \cosh g(\bm z)}{\mathbb{E}_{K+1}\cosh g(\bm z)} - \tanh \tilde g(\bm z) \dfrac{\mathbb{E}_{K+1} \cosh (\bm z) \log \cosh g(\bm z) }{\mathbb{E}_{K+1}\cosh g(\bm z)}\right]. \label{eq:SCEnosquare}
    \end{align}
    \item In order to complete the procedure, we need to expand all the other factors in the expressions. 
    First of all, defined $\tilde{\mathcal{N}}_a$ as 
        \begin{align}
        \tilde{\mathcal{N}}_a = \begin{cases}
            \mathbb{E}_{K+1} \cosh g(\bm z), \quad a=K \\ 
            \mathbb{E}_K \tilde{\mathcal{N}}_K^{\theta_{K-1}}, \quad a=K-1 \\
            \mathbb{E}_{a+1} \tilde {\mathcal{N}}_{a+1}^{\theta_a/\theta_{a+1}}, \quad a=1, \hdots, K-2
        \end{cases}
        \end{align}
    we stress that the expansion of $\mathcal{N}_K=\mathbb{E}_{K+1} \cosh^{\theta_K} g(\bm z)$ is 
    \begin{align}
        \mathcal{N}_K\sim \tilde{\mathcal{N}}_K + (\theta_K-1) \mathbb{E}_{K+1} \cosh g(\bm z) \log \cosh g(\bm z). 
    \end{align}
    As far as the generic $a$ concerns, we focus on $\mathcal{N}_{a}^{\theta_{a-1}/\theta_a}$, whose expansion can be written as 
    \begin{align}
        \begin{cases}
            \mathcal{N}_a^{\theta_{a-1}/\theta_a} \sim \tilde{\mathcal{N}}_a^{\theta_{a-1}/\theta_a} + (\theta_K -1) F_a, \quad a=1, \hdots, K-1, \\
            \mathcal{N}_K^{\theta_{K-1}/\theta_K} \sim \tilde{\mathcal{N}}_K^{\theta_{K-1}} + (\theta_K-1) F_K,
        \end{cases}
        \label{eq:Naexp}
    \end{align} 
    where 
    \begin{align}
        F_a=\begin{cases}
        \dfrac{\theta_{a-1}}{\theta_a} \tilde{\mathcal{N}}_a^{\frac{\theta_{a-1}}{\theta_a}} \dfrac{F_{a+1}}{\tilde{\mathcal{N}}_a}, \quad a=1, \hdots, K-1 \\
            \theta_{K-1} \tilde{\mathcal{N}_K}^{\theta_{K-1}} \left( \dfrac{\mathbb{E}_{K+1} \cosh g(\bm z) \log \cosh g(\bm z)}{\tilde{\mathcal{N}}_K} - \log \tilde{\mathcal{N}}_K\right), \quad a=K.
        \end{cases}
    \end{align}
    Since \eqref{eq:Na1} is valid, in this way we have the expansion either of $\mathcal{N}_{a-1}$ or $\mathcal{N}_a^{\theta_{a-1}/\theta_a}$.
    \item Starting from the innermost expectation of the equation, namely the $E_{K+1}$ one, we made $K-B$ steps exploiting \eqref{eq:expansionmult},\eqref{eq:expansiondiv}, \eqref{eq:Naexp}  when required. 
    \item Now we need to expand the square, so we apply \eqref{eq:expansionsquare} and, after that, continue as in the point 3, until we reach $\mathbb{E}_1$.
\end{enumerate}

To show how the procedure works, we will do the computations for the case of the expansion of 2RSB quenched statistical pressure vs 1RSB one around $\theta_2 = 1$. 

First of all, we will need the expansion of the corresponding terms of \eqref{eq:SCEsquare} and \eqref{eq:SCEnosquare}, which are the $K=2$ case of \eqref{eq:SCEsquare} and \eqref{eq:SCEnosquare}:

\begin{align}
    &\dfrac{\mathbb{E}_{3} \cosh^{\theta_2} g(\bm z) \tanh g(\bm z)}{\mathbb{E}_{3} \cosh^{\theta_2} g(\bm z)} \sim \tanh \tilde g(\bm z) + (\theta_2-1) \notag \\
    &\hspace{4cm}\cdot\left[ \dfrac{\mathbb{E}_{3} \sinh g(\bm z) \log \cosh g(\bm z)}{\mathbb{E}_{3}\cosh g(\bm z)} - \tanh \tilde g(\bm z) \dfrac{\mathbb{E}_{3} \cosh (\bm z) \log \cosh g(\bm z) }{\mathbb{E}_{3}\cosh g(\bm z)}\right] \label{eq:1}
\end{align}
\begin{align}
    \dfrac{\mathbb{E}_{3} \cosh^{\theta_2} g(\bm z) \tanh^2 g(\bm z)}{\mathbb{E}_{3} \cosh^{\theta_2} g(\bm z)} &\sim \dfrac{\mathbb{E}_{3} \sinh g(\bm z) \tanh g(\bm z)}{\mathbb{E}_{3}\cosh g(\bm z)} + (\theta_2-1) \notag \\
    &\cdot \left[ \dfrac{\mathbb{E}_{3} \log \cosh (\bm z)\sinh g(\bm z) \tanh g(\bm z) \mathbb{E}_{3} \cosh g(\bm z) \log \cosh (\bm z)}{\left( \mathbb{E}_{3}\cosh g(\bm z)\right)^2}\right]
    \label{eq:2square}
\end{align}

Let us start from the expansion of $\q_0$ in Eq. \eqref{eq:q0_2RSB}. Since $B=0 \neq 2$, we use \eqref{eq:1}. Then, following the equation, we need to do $K-B=2-0$ steps until the application of the square. 
Now, we compute the corresponding $\mathcal{N}_K^{\theta_{K-1}/\theta_K}$ for $K=2$.
\begin{align}
    &\mathcal{N}_2^{\theta_1/\theta_2} \sim (\mathbb{E}_3 \mathcal{N}_3)^{\theta_1} + (\theta_2 -1 ) \theta_1 \left(\mathbb{E}_3 \mathcal{N}_3 \right)^{\theta_1 }\left[ \dfrac{\mathbb{E}_3 \cosh g(\bm z) \log \cosh g(\bm z) - \mathbb{E}_3 \cosh g(\bm z) \log \mathbb{E}_3 \cosh g(\bm z)}{\mathbb{E}_3 \mathcal{N}_3} \right] \notag \\
    &\hspace{1cm} \sim (\mathbb{E}_3 \cosh g(\bm z))^{\theta_1} \notag \\
    &\hspace{1.8cm}+ (\theta_2 -1 ) \theta_1 \left(\mathbb{E}_3 \cosh g(\bm z) \right)^{\theta_1 }\left[ \dfrac{\mathbb{E}_3 \cosh g(\bm z) \log \cosh g(\bm z) - \mathbb{E}_3 \cosh g(\bm z) \log \mathbb{E}_3 \cosh g(\bm z)}{\mathbb{E}_3 \cosh g(\bm z)} \right]
    \label{eq:N2}
\end{align}

Then, we multiply these two terms together in order to get 
\begin{align}
    &\mathcal{N}_2^{\theta_1/\theta_2} 
 \dfrac{\mathbb{E}_{3} \cosh^{\theta_2} g(\bm z) \tanh g(\bm z)}{\mathbb{E}_{3} \cosh^{\theta_2} g(\bm z)} \sim (\mathbb{E}_3 \cosh g(\bm z))^{\theta_1} \tanh \tilde g(\bm z) \notag \\
 &+ (\theta_2 -1 ) \left\{ \tanh \tilde g(\bm z) (\mathbb{E}_3 \cosh g(\bm z))^{\theta_1} \dfrac{\mathbb{E}_3 \cosh g(\bm z) \log \cosh g(\bm z)}{\mathbb{E}_3 \cosh g(\bm z) } \right.\notag \\
 &\left.- \tanh \tilde g(\bm z)\theta_1 \left(\mathbb{E}_3 \cosh g(\bm z) \right)^{\theta_1 }\left[ \dfrac{\mathbb{E}_3 \cosh g(\bm z) \log \cosh g(\bm z) - \mathbb{E}_3 \cosh g(\bm z) \log \mathbb{E}_3 \cosh g(\bm z)}{\mathbb{E}_3 \cosh g(\bm z)} \right]\right\}, \label{eq:3}
\end{align}
which is the argument of $\mathbb{E}_2$.

Now, we need to expand the ratio between it and the expansion of $\mathcal{N}_1 = \mathbb{E}_2 \left( \mathbb{E}_3 \cosh^{\theta_2} g(\bm z)\right)^{\theta_1/\theta_2}$ which is 
\begin{align}
    &\mathcal{N}_1 \sim\mathbb{E}_2(\mathbb{E}_3 \cosh g(\bm z))^{\theta_1} \notag \\
    &+ (\theta_2-1) \theta_1 \mathbb{E}_2 \left\{ (\mathbb{E}_3 \cosh g(\bm z))^{\theta_1} \left[ \dfrac{\mathbb{E}_3 \cosh g(\bm z) \log \cosh g(\bm z) - \mathbb{E}_3 \cosh g(\bm z) \log \mathbb{E}_3 \cosh g(\bm z)}{\mathbb{E}_3 \cosh g(\bm z)} \right]\right\}.
    \label{eq:2}
\end{align}
We get 
\begin{align}
    &\dfrac{\mathbb{E}_2 \left[\mathcal{N}_2^{\theta_1/\theta_2} 
 \dfrac{\mathbb{E}_{3} \cosh^{\theta_2} g(\bm z) \tanh g(\bm z)}{\mathbb{E}_{3} \cosh^{\theta_2} g(\bm z)}\right]}{\mathcal{N}_1} \sim \dfrac{\mathbb{E}_2 \left[(\mathbb{E}_3 \cosh g(\bm z))^{\theta_1} \tanh \tilde g(\bm z) \right]}{\mathbb{E}_2 \left[\mathbb{E}_3 \cosh g(\bm z) \right]^{\theta_1}} \notag \\
 &\hspace{2cm}+ (\theta_2-1)\left\{ \theta_1\dfrac{\mathbb{E}_2 \left[(\mathbb{E}_3 \cosh g(\bm z))^{\theta_1-1} \mathbb{E}_3 \cosh g(\bm z) \log \cosh(\bm z) \tanh \tilde g(\bm z) \right]}{\mathbb{E}_2 \left[\mathbb{E}_3 \cosh g(\bm z) \right]^{\theta_1}}  \right. \notag \\
 &\hspace{2cm}- \theta_1 \dfrac{\mathbb{E}_2 \left[(\mathbb{E}_3 \cosh g(\bm z))^{\theta_1} \log \mathbb{E}_3 \cosh g(\bm z) \tanh \tilde g(\bm z) \right]}{\mathbb{E}_2 \left[\mathbb{E}_3 \cosh g(\bm z) \right]^{\theta_1}}\notag \\
 &\hspace{2cm}\left. -\dfrac{\mathbb{E}_2 \left[(\mathbb{E}_3 \cosh g(\bm z))^{\theta_1} \tanh \tilde g(\bm z) \right] \mathbb{E}_2 \left[ (\mathbb{E}_3 \cosh g(\bm z))^{\theta_1-1} \mathbb{E}_3 \cosh g(\bm z) \log \cosh g(\bm z) \right]}{\left(\mathbb{E}_2 \left[\mathbb{E}_3 \cosh g(\bm z) \right]^{\theta_1}\right)^2}\right. \notag \\
  &\hspace{2cm}\left. +\dfrac{\mathbb{E}_2 \left[ (\mathbb{E}_3 \cosh g(\bm z))^{\theta_1} \log \mathbb{E}_3 \cosh g(\bm z) \right]}{\left(\mathbb{E}_2 \left[\mathbb{E}_3 \cosh g(\bm z) \right]^{\theta_1}\right)^2}\right\},
\end{align}
which is the argument of the square. 

In the end, we expand it and we get the expansion for $\q_0$ 
\begin{align}
    &\q_0 \sim \mathbb{E}_1 \left\{ \dfrac{\mathbb{E}_2 \left[(\mathbb{E}_3 \cosh g(\bm z))^{\theta_1} \tanh \tilde g(\bm z) \right]}{\mathbb{E}_2 \left[\mathbb{E}_3 \cosh g(\bm z) \right]^{\theta_1}}\right\}^2 \notag \\
    &\hspace{2cm}+ 2(\theta_2-1) \mathbb{E}_1 \left\{ \left[\theta_1\dfrac{\mathbb{E}_2 \left[(\mathbb{E}_3 \cosh g(\bm z))^{\theta_1-1} \mathbb{E}_3 \cosh g(\bm z) \log \cosh(\bm z) \tanh \tilde g(\bm z) \right]}{\mathbb{E}_2 \left[\mathbb{E}_3 \cosh g(\bm z) \right]^{\theta_1}}  \right. \right.\notag \\
 &\hspace{2cm}- \theta_1 \dfrac{\mathbb{E}_2 \left[(\mathbb{E}_3 \cosh g(\bm z))^{\theta_1} \log \mathbb{E}_3 \cosh g(\bm z) \tanh \tilde g(\bm z) \right]}{\mathbb{E}_2 \left[\mathbb{E}_3 \cosh g(\bm z) \right]^{\theta_1}}\notag \\
 &\hspace{2cm}\left. -\dfrac{\mathbb{E}_2 \left[(\mathbb{E}_3 \cosh g(\bm z))^{\theta_1} \tanh \tilde g(\bm z) \right] \mathbb{E}_2 \left[ (\mathbb{E}_3 \cosh g(\bm z))^{\theta_1-1} \mathbb{E}_3 \cosh g(\bm z) \log \cosh g(\bm z) \right]}{\left(\mathbb{E}_2 \left[\mathbb{E}_3 \cosh g(\bm z) \right]^{\theta_1}\right)^2}\right. \notag \\
  &\hspace{2cm}\left.\left. +\dfrac{\mathbb{E}_2 \left[ (\mathbb{E}_3 \cosh g(\bm z))^{\theta_1} \log \mathbb{E}_3 \cosh g(\bm z) \right]}{\left(\mathbb{E}_2 \left[\mathbb{E}_3 \cosh g(\bm z) \right]^{\theta_1}\right)^2}\right] \dfrac{\mathbb{E}_2 \left[(\mathbb{E}_3 \cosh g(\bm z))^{\theta_1} \tanh \tilde g(\bm z) \right]}{\mathbb{E}_2 \left[\mathbb{E}_3 \cosh g(\bm z) \right]^{\theta_1}}\right\} \notag \\
  &\hspace{1cm}\sim \tilde q_0 + C(\q_0, \q_1, \q_2),
  \label{eq:C}
\end{align}
which allows us to make explicit the expression of $C(\q_0, \q_1, \q_2)$.

As far as $\q_1$ concerns, we proceed in the same way as in \eqref{eq:1} - \eqref{eq:3}, but in this case we have to make $K-B=2-1=1$ step until the square, so we get 
\begin{align}
    &\left[
 \dfrac{\mathbb{E}_{3} \cosh^{\theta_2} g(\bm z) \tanh g(\bm z)}{\mathbb{E}_{3} \cosh^{\theta_2} g(\bm z)}\right]^2 \sim \tanh^2 \tilde{g}(\bm z)+ 2(\theta_2-1)\tanh \tilde{g}(\bm z) \notag \\
 &\hspace{3cm}\cdot\left[ \dfrac{\mathbb{E}_{3} \sinh g(\bm z) \log \cosh g(\bm z)}{\mathbb{E}_{3}\cosh g(\bm z)} - \tanh \tilde g(\bm z) \dfrac{\mathbb{E}_{3} \cosh (\bm z) \log \cosh g(\bm z) }{\mathbb{E}_{3}\cosh g(\bm z)}\right].
 \label{eq:square2}
\end{align}
Then, we expand the multiplication between \eqref{eq:square2} and $\mathcal{N}_2^{\theta_1/\theta_2}$ as in \eqref{eq:N2}: 
\begin{align}
    &\mathcal{N}_2^{\theta_1/\theta_2}\left[
 \dfrac{\mathbb{E}_{3} \cosh^{\theta_2} g(\bm z) \tanh g(\bm z)}{\mathbb{E}_{3} \cosh^{\theta_2} g(\bm z)}\right]^2 \sim (\mathbb{E}_3 \mathcal{N}_3)^{\theta_1} \tanh^2 \tilde{g}(\bm z) + (\theta_2-1)\notag \\
 &\hspace{1cm}\left[ \theta_1 \tanh^2 \tilde{g}(\bm z)\left(\mathbb{E}_3 \cosh g(\bm z) \right)^{\theta_1 }\left[ \dfrac{\mathbb{E}_3 \cosh g(\bm z) \log \cosh g(\bm z) - \mathbb{E}_3 \cosh g(\bm z) \log \mathbb{E}_3 \cosh g(\bm z)}{\mathbb{E}_3 \cosh g(\bm z)} \right] \right.\notag \\
 &\hspace{1cm}\left.+ (\mathbb{E}_3 \mathcal{N}_3)^{\theta_1}\left[ \dfrac{\mathbb{E}_{3} \sinh g(\bm z) \log \cosh g(\bm z)}{\mathbb{E}_{3}\cosh g(\bm z)} - \tanh \tilde g(\bm z) \dfrac{\mathbb{E}_{3} \cosh (\bm z) \log \cosh g(\bm z) }{\mathbb{E}_{3}\cosh g(\bm z)}\right]\right], 
\end{align}
which is the argument of $\mathbb{E}_2$. 

In the end, we just need to make the expansion of the ratio between it and \eqref{eq:2} to get 

\begin{align}
    \q_1\sim& \mathbb{E}_1 \left\{ \dfrac{\mathbb{E}_2 \left[\cosh^{\theta_1} \tilde g(\bm z)\tanh^2 \tilde g( \bm z)\right] } {\mathbb{E}_2 \cosh^{\theta_1} \tilde g(\bm z)}\right\} + (\theta_2-1) \mathbb{E}_1 \left\{ \dfrac{1}{\tilde{N}_1^2} \left[ \mathcal{N}_1 \mathbb{E}_2\left[ \theta_1 \tanh^2 \tilde{g}(\bm z)\left(\mathbb{E}_3 \cosh g(\bm z) \right)^{\theta_1 }\right. \right. \right. \notag \\
    &\left[ \dfrac{\mathbb{E}_3 \cosh g(\bm z) \log \cosh g(\bm z) - \mathbb{E}_3 \cosh g(\bm z) \log \mathbb{E}_3 \cosh g(\bm z)}{\mathbb{E}_3 \cosh g(\bm z)} \right] \notag \\
    &\left. \left. \left. + (\mathbb{E}_3 \mathcal{N}_3)^{\theta_1}\left[ \dfrac{\mathbb{E}_{3} \sinh g(\bm z) \log \cosh g(\bm z)}{\mathbb{E}_{3}\cosh g(\bm z)} - \tanh \tilde g(\bm z) \dfrac{\mathbb{E}_{3} \cosh (\bm z) \log \cosh g(\bm z) }{\mathbb{E}_{3}\cosh g(\bm z)}\right]\right]   \right.\right.\notag \\
    &- \mathbb{E}_2 \left[\cosh^{\theta_1} \tilde g(\bm z)\tanh^2 \tilde g( \bm z)\right] \theta_1\mathbb{E}_2 \left\{ (\mathbb{E}_3 \cosh g(\bm z))^{\theta_1} \right.\notag \\
    &\left. \left.\left. \cdot\left[ \dfrac{\mathbb{E}_3 \cosh g(\bm z) \log \cosh g(\bm z) - \mathbb{E}_3 \cosh g(\bm z) \log \mathbb{E}_3 \cosh g(\bm z)}{\mathbb{E}_3 \cosh g(\bm z)} \right]\right\} \right] \right\} \\
    \sim& \tilde q_1 + (\theta_2-1) B(\q_0, \q_1, \q_2), 
    \label{eq:B}
\end{align}
which gives us the expression of $B(\q_0, \q_1, \q_2)$.

Finally, for $\q_2$ we use \eqref{eq:2square} 
and we do all the computations we have done before in order to get 
\begin{align}
    \q_2 \sim& \mathbb{E}_1 \left\{ \dfrac{\mathbb{E}_2 \cosh^{\theta_1} \tilde g( \q_1, \q_0)\dfrac{\mathbb{E}_3 \cosh g (\q_0, \q_1, \q_2) \tanh^2 g (\q_0, \q_1, \q_2, \mm)}{\mathbb{E}_3 \cosh^{\theta_1} g (\q_0, \q_1, \q_2)}} {\mathbb{E}_2 \cosh^{\theta_1} \tilde g(\mm, \q_1, \q_0)}\right\} \notag \\
    &+ (\theta_2-1) \mathbb{E}_1 \left\{ \dfrac{\mathbb{E}_2 \left[  \theta_1 \left( \mathbb{E}_3 \cosh g \right)^{-1}\mathbb{E}_3 \cosh g(\bm z) \log \cosh g(\bm z) \mathbb{E}_3 \cosh g(\bm z) \tanh^2 g(\bm z) \right]}{\mathbb{E}_2 \left( \mathbb{E}_3 \cosh g \right)^{\theta_1}} \right. \notag \\
    &\left.- \dfrac{\mathbb{E}_2 \left[  \theta_1 \log \mathbb{E}_3 \cosh g(\bm z) \mathbb{E}_3 \cosh g(\bm z) \tanh^2 g(\bm z)\right]}{\mathbb{E}_2 \left( \mathbb{E}_3 \cosh g \right)^{\theta_1}} \right. \notag \\
    &\left. - \dfrac{\mathbb{E}_2 \left[  \mathbb{E}_3 \cosh g(\bm z) \tanh^2 g(\bm z) \right]\mathbb{E}_2 \left[ \left( \mathbb{E}_3 \cosh g \right)^{\theta_1-1} \mathbb{E}_3 \cosh g(\bm z) \log \cosh g(\bm z) \right]}{\mathbb{E}_2 \left( \mathbb{E}_3 \cosh g \right)^{\theta_1}}\right. \notag \\
    &\left. +\dfrac{\mathbb{E}_2 \left[  \mathbb{E}_3 \cosh g(\bm z) \tanh^2 g(\bm z) \right]\mathbb{E}_2 \left[ \left( \mathbb{E}_3 \cosh g \right)^{\theta_1}\log \mathbb{E}_3 \cosh g(\bm z) \right]}{\left(\mathbb{E}_2 \left( \mathbb{E}_3 \cosh g \right)^{\theta_1}\right)^2}\right\} \\
    \sim& \tilde q_2 + (\theta_2-1) A( \q_0, \q_1, \q_2), 
    \label{eq:A}
\end{align}
which gives us the expression of $A(\q_0, \q_1, \q_2)$.

Therefore, in this Section we have shown the computations of the correction terms in $\theta_K \to 1$ limit. The iterative procedure will be the same in any other limit cases. 
\end{document}